\documentclass[traditabstract]{aa} % for the abstract without structuration 
\usepackage{epsfig}
\usepackage{txfonts}
\newcommand{\rscnc}{RS\,Cnc}
\newcommand{\COone}{\mbox{CO1-0}}
\newcommand{\COtwo}{\mbox{CO2-1}}
\newcommand{\Lsol}{L$_{\odot}$}
\newcommand{\Rsol}{R$_{\odot}$}
\newcommand{\Msol}{M$_{\odot}$}
\newcommand{\HI}{H\,{\sc {i}}~}

\newcommand{\Msold}{M$_{\odot}$\,yr$^{-1}$}

\newcommand{\Vstar}{V$_{\star}$}
\newcommand{\Vlsr}{V$_{\rm lsr}$}

\newcommand{\Teff}{T$_{\rm eff}$}
\newcommand{\kms}{km\,s$^{-1}$}

\newcommand{\lsim}{~\rlap{$<$}{\lower 1.0ex\hbox{$\sim$}}}
\newcommand{\gsim}{~\rlap{$>$}{\lower 1.0ex\hbox{$\sim$}}}
\newcommand{\as}[2]{$#1''\,\hspace{-1.7mm}.\hspace{.0mm}#2$}
\newcommand{\am}[2]{$#1'\,\hspace{-1.7mm}.\hspace{.0mm}#2$}

\begin{document}
 \title{The multi-scale environment of RS Cnc from CO and \HI observations\thanks{Based on observations carried out with the IRAM Plateau de Bure Interferometer and the IRAM 30-m telescope. IRAM is supported by INSU/CNRS (France), MPG (Germany) and IGN (Spain).}}

   \author{D. T. Hoai\inst{1,2},  
           L. D. Matthews\inst{3},
	   J. M. Winters\inst{4},
           P. T. Nhung\inst{1,2},
	   E. G\'erard\inst{5},
           Y. Libert\inst{1,4},
	   \and T. Le\,Bertre\inst{1}
          }

   \institute{LERMA, UMR 8112, CNRS \& Observatoire de Paris, 
	      61 av. de l'Observatoire, F-75014 Paris, France 
%              \email{name.surname@obspm.fr}
         \and
              VATLY/INST, 179 Hoang Quoc Viet, Cau Giay, Ha Noi, Vietnam
         \and
              MIT Haystack Observatory, Off Route 40,
              Westford, MA 01886, USA
         \and
              IRAM, 300 rue de la Piscine, Domaine Universitaire, 
	      F-38406 St. Martin d'H\`eres, France
         \and
	      GEPI, UMR 8111, CNRS \& Observatoire de Paris, 
	      5 Place J. Janssen, F-92195 Meudon Cedex, France
             }

%   \date{\today}
   \date{Received January 29, 2014; accepted March 07, 2014}

   \titlerunning{The multi-scale environment of RS Cnc}
   \authorrunning{D. T. Hoai, L. D. Matthews, J. M. Winters, et al.}

% \abstract{}{}{}{}{} 
% 5 {} token are mandatory
 
  \abstract
  % context heading (optional)
{We present a detailed study of the circumstellar gas distribution and 
kinematics of the semi-regular variable star RS Cnc on spatial scales ranging 
from $\sim 1''$ ($\sim$150~AU) to $\sim 6'$ ($\sim$0.25 pc). 
Our study utilizes 
new CO1-0 data from the Plateau-de-Bure Interferometer and new \HI 21-cm line 
observations from the Jansky Very Large Array (JVLA), in combination with 
previous observations. New modeling of CO1-0 and CO2-1 imaging observations 
leads to a revised characterization of RS Cnc's previously identified 
axisymmetric molecular outflow. Rather than a simple disk-outflow picture, 
we find that a gradient in velocity as a function of latitude is needed to fit 
the spatially resolved spectra, and in our preferred model, the density and 
the velocity vary smoothly from the equatorial plane to the polar axis. 
In terms of density, the source appears quasi-spherical, whereas in terms of 
velocity the source is axi-symmetric with a low expansion velocity in the 
equatorial plane and faster outflows in the polar directions. 
The flux of matter is also larger in the polar directions than 
in the equatorial plane. An implication of 
our model is that the stellar wind is still accelerated at radii larger than 
a few hundred AU, well beyond the radius where the terminal velocity is 
thought to be reached in an asymptotic giant branch star. The JVLA \HI data 
show the previously detected head-tail morphology, but also supply additional 
detail about the atomic gas distribution and kinematics. We confirm that the 
`head' seen in \HI is elongated in a direction consistent with the polar axis 
of the molecular outflow, suggesting that we are tracing an extension of the 
molecular outflow well beyond the molecular dissociation radius (up to 
$\sim$0.05 pc). The $6'$-long \HI `tail' is oriented at a PA of 
305$^{\circ}$, consistent with the space motion of the star. The tail is 
resolved into several clumps that may result from hydrodynamic effects linked 
to the interaction with the local interstellar medium. We measure a total mass 
of atomic hydrogen $M_{\rm HI}\approx 0.0055 M_{\odot}$ and estimate 
a lower limit to the timescale for the formation of the tail to be 
$\sim6.4\times10^4$ years.}

   \keywords{Stars: AGB and post-AGB  --
                {\it (Stars:)} circumstellar matter  --
                Stars: individual: RS\,Cnc  --
                Stars: mass-loss  -- 
                radio lines: stars.
               }

   \maketitle
%
%________________________________________________________________

\section{Introduction}

Asymptotic Giant Branch (AGB) stars are undergoing mass loss at a high rate.  
One of the best tracers of AGB outflows are the rotational lines of carbon 
monoxide (CO). From the modeling of the line profiles it has been possible 
to derive reliable expansion velocities and mass loss rates (Ramstedt et al. 
2008). In addition imaging at high spatial resolution allows us to describe 
the geometry and the kinematics of these outflows in the inner 
circumstellar regions where the winds emerge and where their main
characteristics get established (Neri et al. 1998). 

High quality observations of CO line emissions at high spectral resolution 
have shown that some profiles are composite, with a narrow component 
superimposed on a broader one, revealing the presence of two winds with 
different expansion velocities (Knapp et al. 1998, Winters et al. 2003).
Using high spatial resolution data obtained in the CO1-0 and 2-1 lines,  
Libert et al. (2010) have suggested that the composite line-profiles of 
the semi-regular AGB star 
\rscnc ~probably originate from an axi-symmetrical geometry with a slowly 
expanding equatorial disk and a faster perpendicular bipolar outflow. 
Other cases of AGB stars with axi-symmetrical expanding shells have been 
identified in a CO mapping survey of AGB stars by Castro-Carrizo et al. (2010).
It shows that the axi-symmetry which is often observed in post-AGB stars 
(e.g. Sahai et al. 2007) may develop earlier when the stars are still on 
the AGB.

Although extremely useful, CO as a tracer is limited to the inner parts 
of the circumstellar shells because, at a distance of typically $\sim$ 
10$^{17}$\,cm, it is photo-dissociated by the interstellar radiation field 
(ISRF). At larger distances, it is necessary to use other tracers, such as 
dust or atomic species. The \HI line at 21\,cm has proved to be an excellent 
spatio-kinematic tracer of the external regions of circumstellar shells 
(e.g. G\'erard \& Le~Bertre 2006, Matthews \& Reid 2007). In particular, 
the \HI map of \rscnc ~presented by Matthews \& Reid shows a 6$'$-long tail, 
in a direction opposite to the space motion of the central star, and clearly 
different from that of the bipolar flow observed in CO at shorter distances 
(2--10$''$) by Libert et al. (2010). In such a case the shaping mechanism 
is thought to be due to the motion of the star relative to the local 
interstellar medium (Libert et al. 2008, Matthews et al. 2013). 

Thus \rscnc ~is an ideal target to study, in the same source, 
the two main effects that are expected to shape circumstellar environments, 
and to evaluate their respective roles. 
In this paper, we revisit \rscnc ~with new high spatial resolution data 
obtained in CO at 2.6 mm and in \HI at 21 cm. 
Our goal is to combine observations in these two complementary tracers, 
in order to describe the spatio-kinematic structure of the circumstellar shell 
from its center to the interstellar medium (ISM). The stellar properties of 
\rscnc ~are summarized in Table~\ref{basicdata}. 

Until recently, a distance of 122 pc was adopted from the parallax 
measured using Hipparcos (Perryman et al. 1997). However, new analyses of 
the Hipparcos data led to somewhat larger estimates of the distance, 
129$_{-16}^{+16}$  pc (Famaey et al. 2005) 
and 143$_{-10}^{+12}$ \,pc (van Leeuwen 2007). 
In the present work, we adopt the improved 
values of the parallax and proper motions by van Leeuwen, and 
scale the published results with the new estimate of the distance. 
We also adopt the peculiar solar motion from Sch\"onrich et al. (2010). 
RS\,Cnc is an S-type star (CSS\,589, in Stephenson's (1984) catalogue) in the 
Thermally-Pulsing AGB phase of evolution: Lebzelter \& Hron (1999) 
reported the presence of Tc lines in its spectrum. 

\begin{table}
%\centering
\caption{Properties of \rscnc.}
\begin{tabular}{lll}
\hline
parameter             &    value                  & ref.\\
\hline
distance              &    143 pc                 & 1 \\
MK Spectral type      &   M6eIb-II(S)             & 2 \\
variability type      &     SRc:                  & 2 \\
pulsation periods     & 122 and 248 days          & 3 \\
effective temperature &     3226 K                & 4 \\
radius                &  225 \Rsol                & 4 \\
luminosity            &  4945 \Lsol               & 4 \\
LSR radial velocity (\Vstar) &    6.75 \kms       & 5 \\
expansion velocity    &   2.4/8.0 \kms            & 5 \\
mass loss rate        &  1.7 10$^{-7}$ \Msold     & 6 \\
3D space velocity; PA &  15 \kms; 155$^\circ$     & this work \\
\hline
\end{tabular}\\
{\bf References in the table:\\} 
(1) Hipparcos (van Leeuwen 2007)\\
(2) GCVS (General Catalogue of Variable Stars)\\ 
(3) Adelman \& Dennis (2005)\\
(4) Dumm \& Schild (1998)\\
(5) Libert et al. (2010)\\
(6) Knapp et al. (1998)\\

\label{basicdata}
\end{table}

%__________________________________________________________________

\section[]{CO observations}
\label{COdata}
 
\subsection{summary of previous data}
\label{olddata}

RS Cnc was imaged in the CO1-0 and 2-1 lines by Libert et al. (2010). 
An on-the-fly (OTF) map covering a region of 100$''\times$100$''$ with 
steps of 4$''$ in RA and 5$''$ in Dec was obtained at the IRAM 30-m telescope. 
Interferometric data were  obtained with the Plateau-de-Bure 
Interferometer (PdBI) in three configurations, B, C and D, 
i.e. with baselines ranging from 24-m to 330-m. All sets of observations 
were obtained with a spectral resolution corresponding to 0.1 \kms. 
The data from the 30-m 
telescope and the PdBI were merged and images in 1-0, with a 
field of view (fov) of 44$''$ and a spatial resolution of $\sim$2.3$''$, 
and in 2-1, with a fov of 22$''$ and a spatial resolution of $\sim$1.2$''$, 
were produced. Libert et al. presented the corresponding channel maps 
with a spectral resolution of 0.4 \kms. 

These maps show clearly that the broad and narrow spectral components 
reported by Knapp et al. (1998) originate from two different regions that 
Libert et al. (2010) described as a slowly expanding ($\sim$ 2\,\kms) 
equatorial disk/waist and a faster ($\sim$\,8\,\kms) bipolar outflow. 
Libert et al. estimated that the polar axis lies at an inclination of
$\sim45^{\circ}$ with respect to the plane of the sky and is projected
almost north-south, along a position angle (PA) of $\sim10^{\circ}$.

\subsection{new data}
\label{newdata}

In January and February 2011, we obtained new data in
the \COone ~line with the PdBI array in configurations A and B,
increasing the baseline coverage up to 760\,m.

The new data were obtained in dual polarizations and covered a bandwidth 
of 3.6\,GHz centered at 115.271\,GHz, the nominal frequency of the CO1-0 
line. Two units of the  narrow band correlator were set up to
cover the CO line with a spectral resolution of 39\,kHz over a
bandwidth of 20\,MHz, and the adjacent continuum was observed by the wideband
correlator WideX (Wideband Express) with a channel spacing of 1.95\,MHz.

These observations resulted in 12\,h of on-source integration time with the 6
element array and reach a 1$\sigma$ thermal noise level of 12\,mJy/beam 
in 0.2\,km\,s$^{-1}$ channels. The synthesized beam is
0.92$''\times$ 0.78$''$ at a position angle PA$ = 62^{\circ}$.

\subsection{continuum}
\label{continuum}

We used the new WideX data to produce a continuum image of RS Cnc at 115 GHz. 
The data were integrated over a 2 GHz band, excluding the high frequency 
portion of the band that is affected by atmospheric absorption.
A single point source is clearly detected. The source is unresolved, and 
there is no evidence of a companion. It has a flux density of 5.4$\pm$0.3 mJy, 
which is consistent with the flux density reported by Libert et al. (2010).
It is slightly offset south-west with respect to the center of phase 
because we used the coordinates at epoch 2000.0 from Hipparcos. 
The offset (--0.15$''$ in RA 
and --0.37$''$ in Dec) is consistent with the proper motion reported by 
Hipparcos ($-$11.12\,mas\,yr$ ^{-1}$ in RA and --33.42 mas yr$ ^{-1}$ in Dec, 
van Leeuwen 2007). 

\begin{figure}
\centering
\epsfig{figure=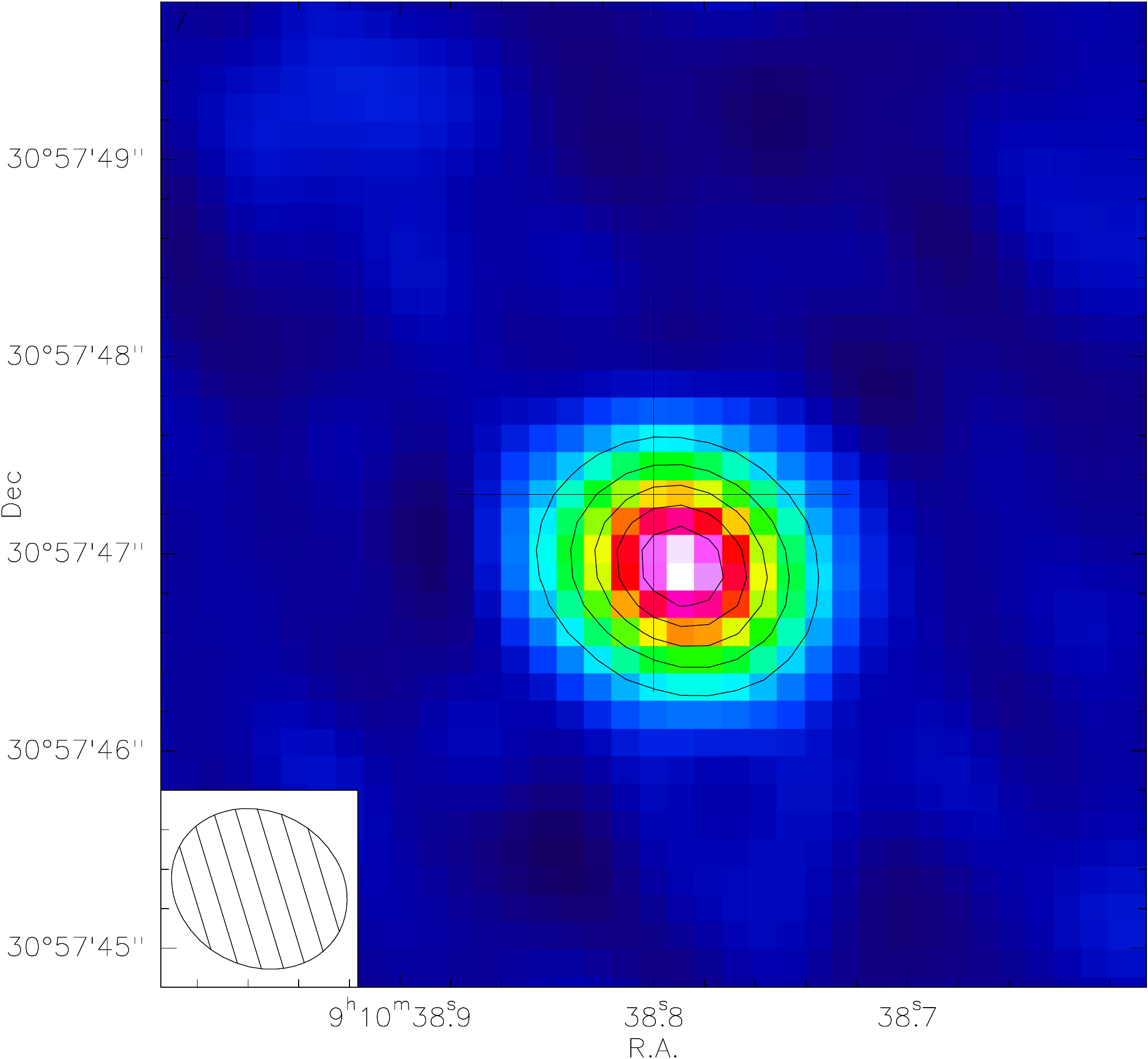,angle=0,width=8.0cm}
\caption{Continuum map at 115 GHz of RS Cnc (A+B configuration data obtained 
in 2011). The cross corresponds to the 2000.0 position of the star  
(RA 09:10:38.800, Dec 30:57:47.30). The contour levels 
are separated by steps of 0.90mJy/beam ($\equiv$\,20$\sigma$). 
The beam is 0.92$''\times0.78''$ (PA = 62$^{\circ}$).}
  \label{continuummap10}
\end{figure}

\subsection{channel maps in CO1-0}
\label{channelmaps}

\begin{figure}
\centering
\epsfig{figure=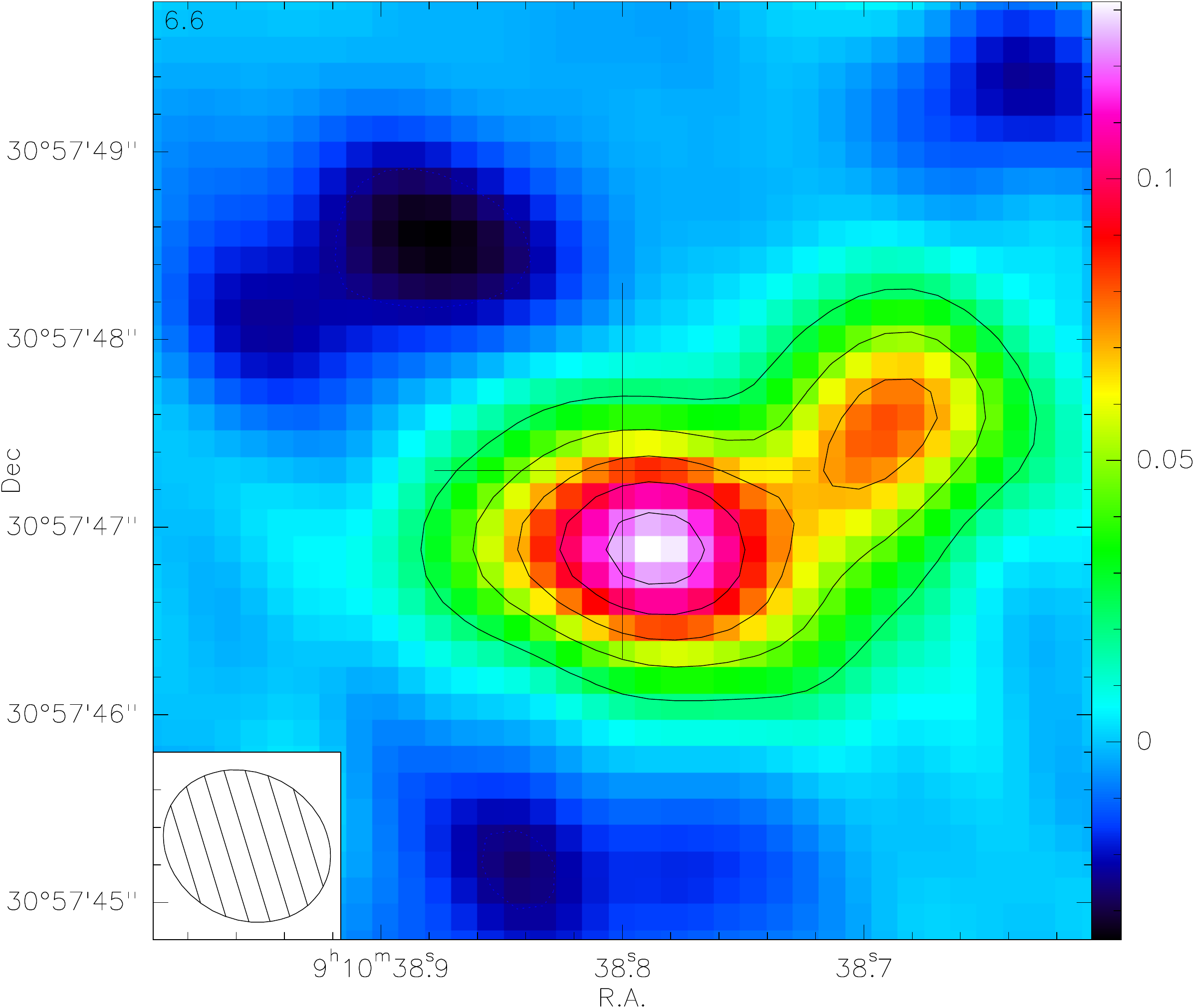,angle=0,width=8.8cm}
\caption{Continuum subtracted CO1-0 channel map at 6.6 \kms ~(A+B 
configuration data obtained in 2011). The contour levels are separated by 
steps of 24 mJy/beam ($\equiv\,2\sigma$). The negative contours are in 
dotted lines.}
  \label{6p6kmpersec-channelmap10}
\end{figure}

The channel maps obtained in 2011 across the CO1-0 line are dominated by 
a compact central source. However, at 6.6\,\kms 
~(Fig.~\ref{6p6kmpersec-channelmap10}), we observe 
a companion source at $\sim$1$''$ west-north-west (PA$\sim$300$^\circ$). 
This source is detected from 5.8 to 6.8\,\kms, but not outside this range.
In the channel map at 6.6\,\kms, in which it is best separated, 
it has an integrated flux of 93$\pm$6 mJy, as compared to 300$\pm$20 mJy 
for the central source. This secondary source is clearly not at the origin 
of the bipolar outflow, which is aligned on the central source. Within errors, 
the central source coincides with the continuum source discussed in the 
previous section.

\subsection{merging (CO1-0)}
\label{merging}

Finally, the new data (observed in the extended A and B
configurations) were merged with the old ones already presented by
Libert et al. (2010).  These were obtained in the previous B, C, and
D configurations and combined with short spacing observations obtained
on the IRAM 30\,m telescope. The final combined data set now covers 
a spectral bandwidth of 580\,MHz, resampled to a spectral resolution of
0.2\,km\,s$^{-1}$. The 1 $\sigma$ thermal noise in the combined data cube 
is 8.7\,mJy/beam (for a channel width of 0.2 \kms) and the synthesized beam 
is 1.15$''\,\times$0.96$''$ at PA\,=\,67$^\circ$, similar to the resolution
already obtained on CO2-1 (Libert et al. 2010). 

\begin{figure*}
\centering
\epsfig{figure=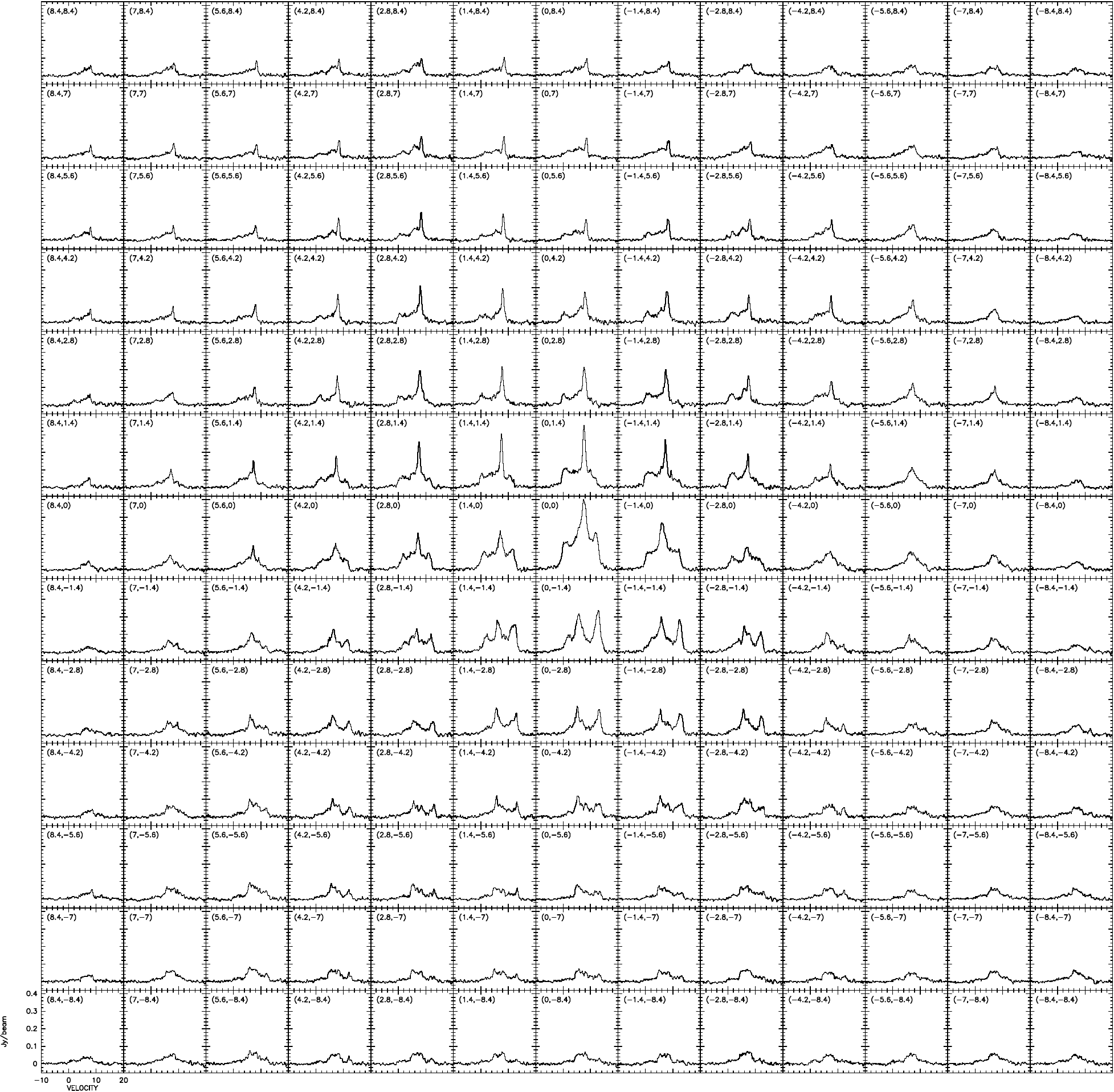,angle=0,width=18.cm}
\caption{CO1-0 spectral map of RS Cnc obtained with a restoring beam 
of 1.2$''$. The offsets with respect to the center of phase are given in 
the upper left corner of each panel (step=1.4$''$). North is up, 
and east is to the left.}
  \label{spectralmapCO10}
\end{figure*}

\begin{figure}
\centering
\epsfig{figure=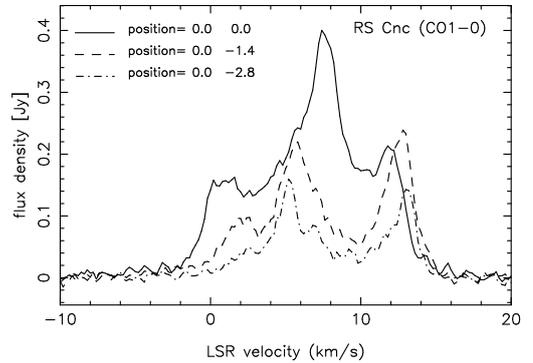,angle=0,width=8.0cm}
\caption{Spectra obtained at different positions (centre, 1.4$''$ south, 
2.8$''$ south) in a beam of 1.2$''$ (cf. Fig.~\ref{spectralmapCO10}). }
  \label{triple-CO10}
\end{figure}

In Fig.~\ref{spectralmapCO10}, we present the resulting spectral map that 
we have derived by using a circular restoring beam with a gaussian profile of 
FWHM\,=\,1.2$''$. 
The line profiles are composed of three components 
whose relative intensities depend on the position in the map. One notes 
also that the two extreme components, at $\sim$\,2\,\kms ~and $\sim$12 \kms, 
tend to deviate more from the central component as the distance to the central 
source increases. In Fig.~\ref{triple-CO10}, three spectra 
obtained in the south of the central position are overlaid. One notes  
a shift in velocity of the red peaks, which correspond to the southern polar 
outflow at $\sim$12-14 \kms) relative to the central peaks at $\sim$5-8 \kms, 
which correspond to the emission close to the equatorial plane, 
with distance to the central source.

\section[]{CO model}
\label{COmodel}

\subsection{description}
\label{description} 

In order to constrain the spatio-kinematic structure of \rscnc, we have 
constructed a model of CO emission adapted to any geometry and based on a 
code already developed by Gardan et al. (2006). A ray-tracing approach, 
taking into account the velocity-dependent emission and absorption 
of each element along a line of sight, allows us to reconstruct the flux 
obtained, within an arbitrary beam, from a source which has 
an arbitrary geometry. 
The density, the excitation temperature and the velocity are defined at each 
point of the circumstellar shell. The code can then produce synthetic 
spectral maps that can be compared to the observed ones. 

\begin{figure}
\centering
\epsfig{figure=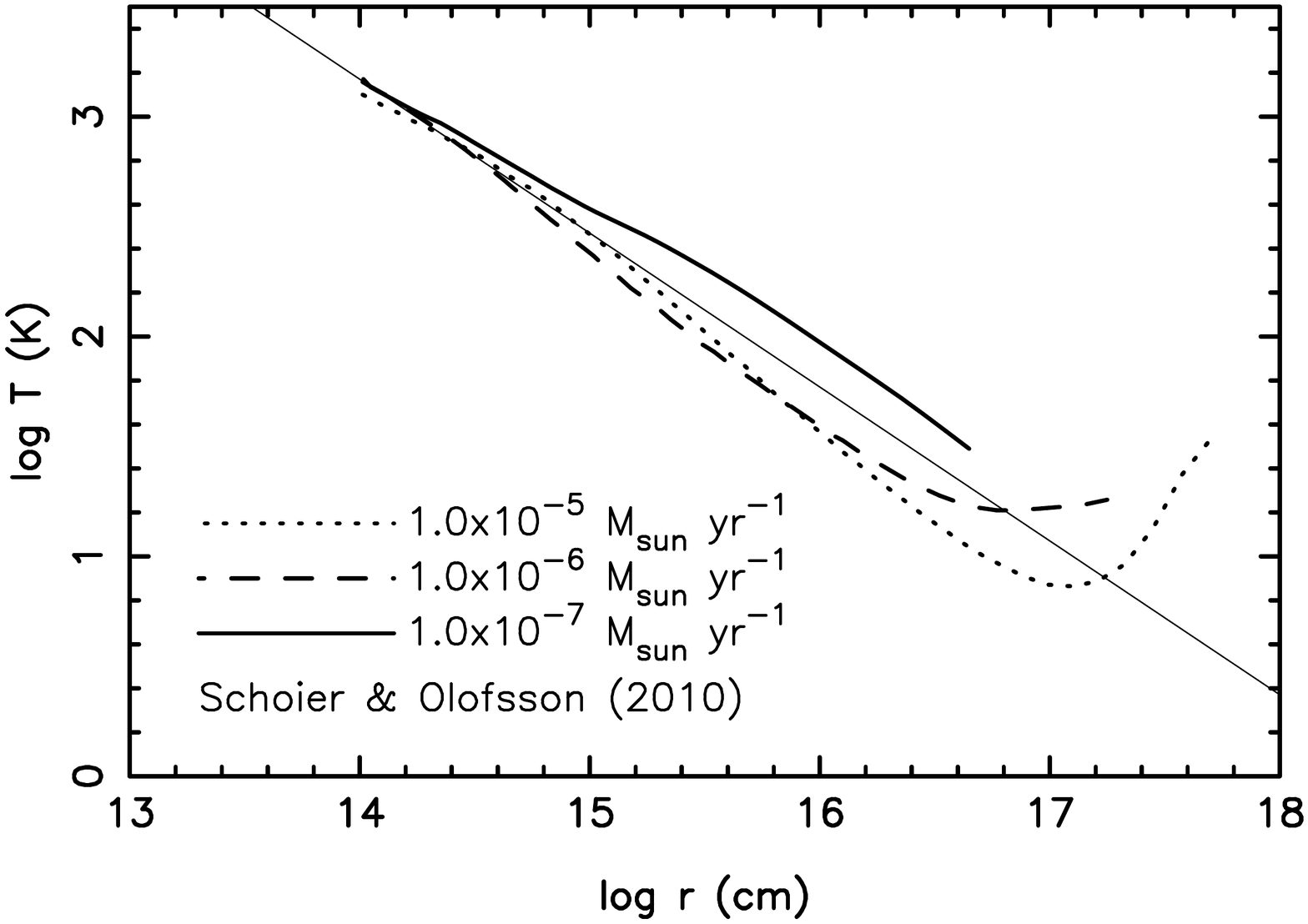,angle=0,width=8.0cm}
\caption{Temperature profiles for different mass loss rates 
from Sch{\"o}ier \& Olofsson (2010, private communication). 
The thin line represents the temperature profile adopted in this work.}
  \label{tempRSCnc}
\end{figure}

The populations of the rotational levels of the CO molecules are calculated 
assuming local thermodynamic equilibrium. The temperature profile is assumed 
to vary as r$^{-0.7}$, where r is the distance from the center of the star, 
and is scaled to models kindly provided by Sch\"oier \& 
Olofsson (2010, private communication; Fig.~\ref{tempRSCnc}). The latter 
profiles were obtained using the radiative transfer code developed 
in spherical geometry by Sch\"oier \& Olofsson (2001). The same temperatures 
are adopted to calculate for each element the thermal Doppler broadening, 
assuming a Maxwellian distribution of the velocities.

\subsection{application to RS Cnc}
\label{application}

Following Libert et al. (2010) the source is defined by an equatorial plane  
and a polar axis. It is thus axi-symmetric and we need only two angles,  
which for instance define the orientation of the polar axis. Hereby, 
for simplicity we will use the angle of inclination of this axis over 
the plane of the sky (AI), and the position angle of the projection on 
the plane of the sky of this axis of symmetry (PA). 

We assume that the velocities are radial and that the outflows are stationary. 
Thus, the product v$\times$n$\times$r$^2$ (where v is the velocity, n the 
density, and r the distance to the centre) is kept constant along every 
radial direction. In order to account for the velocity gradient observed 
in the line profiles (Fig.\,\ref{triple-CO10}), we adopt a dependence of the 
velocity in r$^{\alpha}$, $\alpha$ being the logarithmic velocity gradient 
(Nguyen-Q-Rieu et al. 1979).

We adopt a stellar CO/H abundance ratio of 4.0\,10$^{-4}$ (all carbon in CO, 
Smith \& Lambert 1986). Then we take a dependence of the CO abundance ratio 
with r from the photo-dissociation model of Mamon et al. (1988) 
for a mass loss rate of 1.0$\times10^{-7}$ \Msold ~(see below). 
The external limit is set at 20$''$ ($\sim$\,4.3\,10$^{16}$\,cm).  
In addition, we assume an He/H abundance ratio of 1/9.
The star is offset from the center of the map by the amount measured on 
the continuum map (Fig.~\ref{continuummap10}). Finally, we adopt 
a stellar radial velocity \Vlsr\,=\,6.75\,\kms ~(cf. Table\,\ref{basicdata}). 

\begin{figure}
\centering
\epsfig{figure=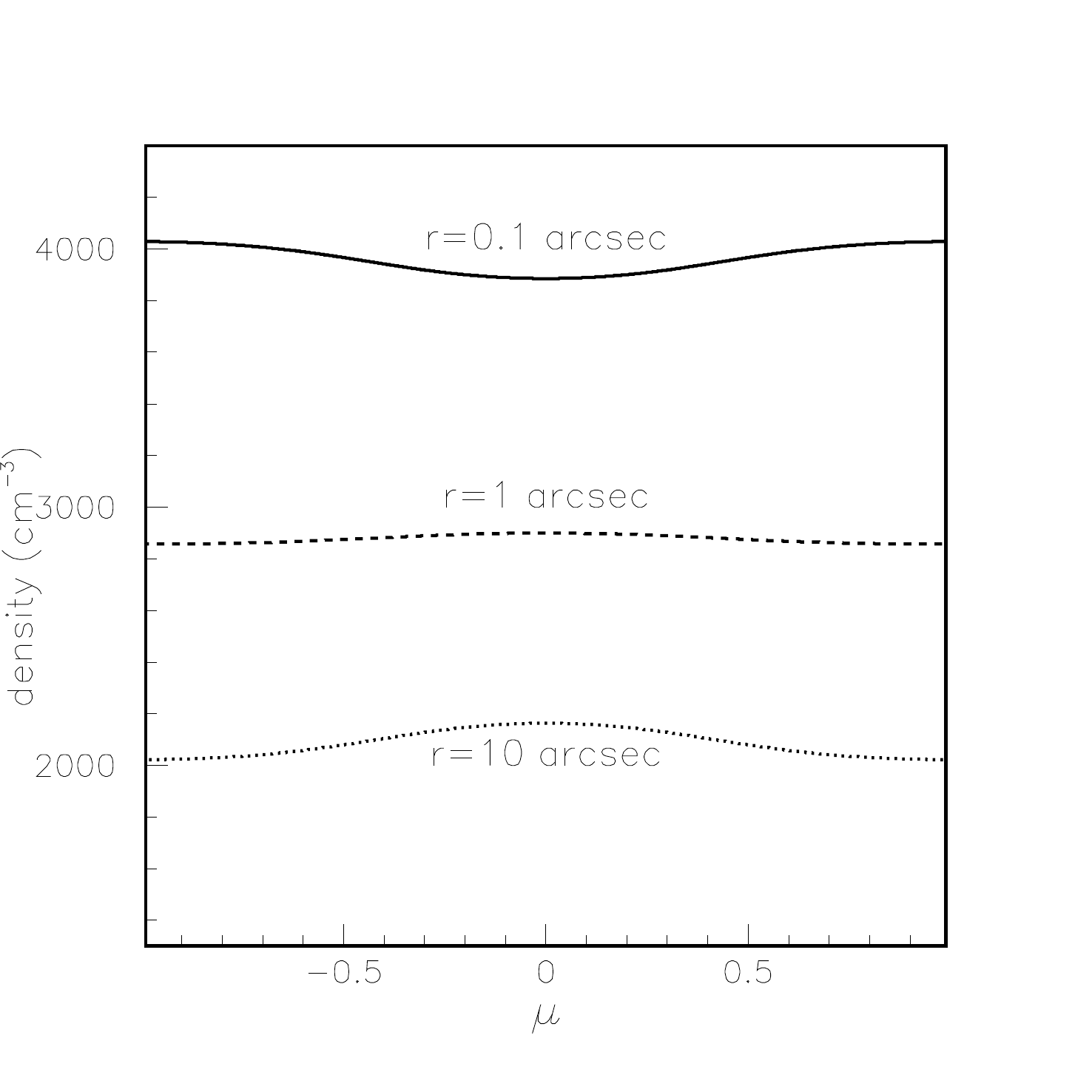,angle=0,width=8.0cm}
\caption{The density profiles (in hydrogen atom number) 
in our preferred model as a function of $\mu$ 
($\equiv$\,sin\,$\theta$, $\theta$ being the angle with respect to the 
equator) and for various distances from the central star. The profile for 
r=1 arcsec (r= 0.1 arcsec) is scaled by a factor 1/100 (1/10000, 
respectively).}
  \label{profd}
\end{figure}

\begin{figure}
\centering
\epsfig{figure=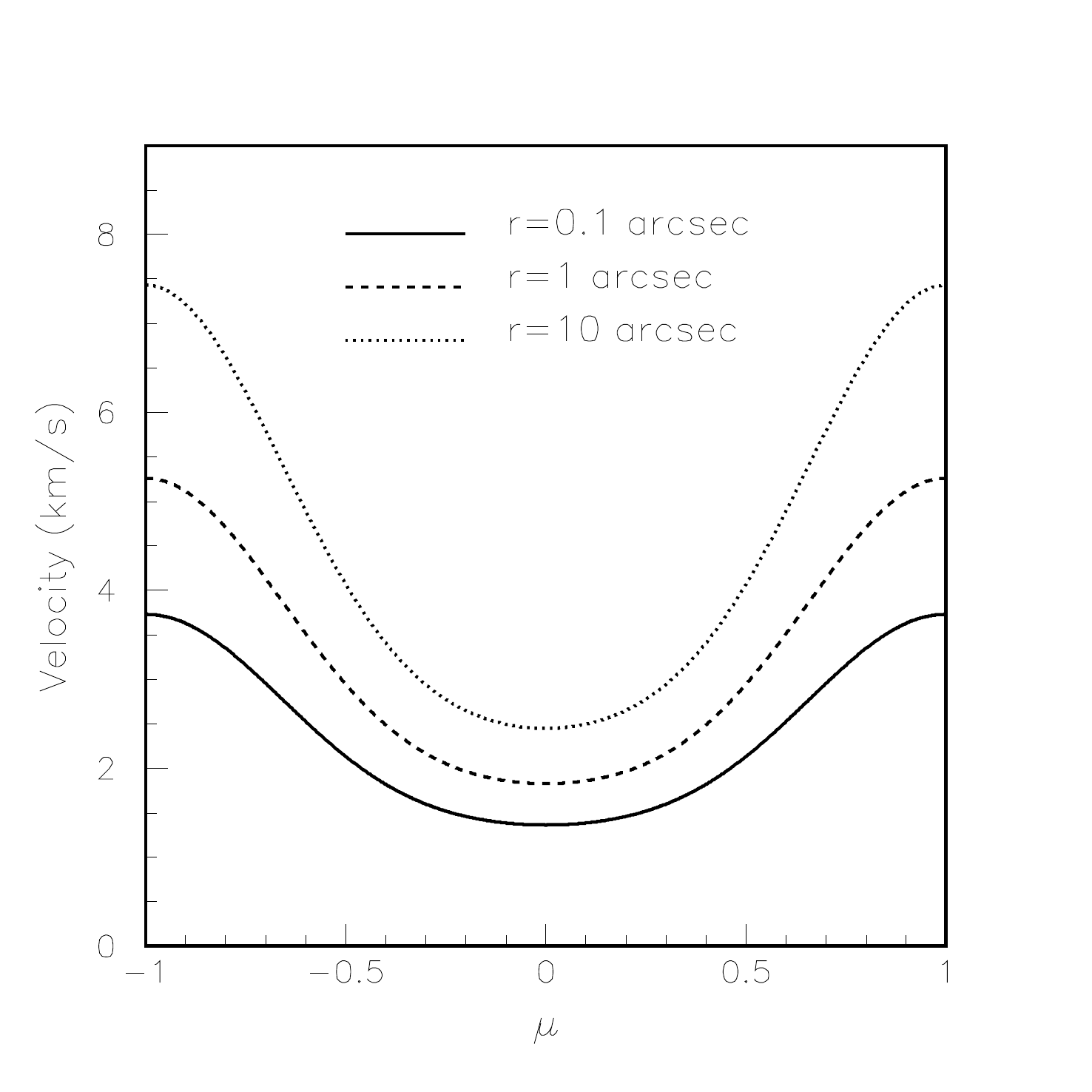,angle=0,width=8.0cm}
\caption{Same as in Fig.~\ref{profd}, but for the velocity.}
  \label{profv}
\end{figure}

\begin{figure}
\centering
\epsfig{figure=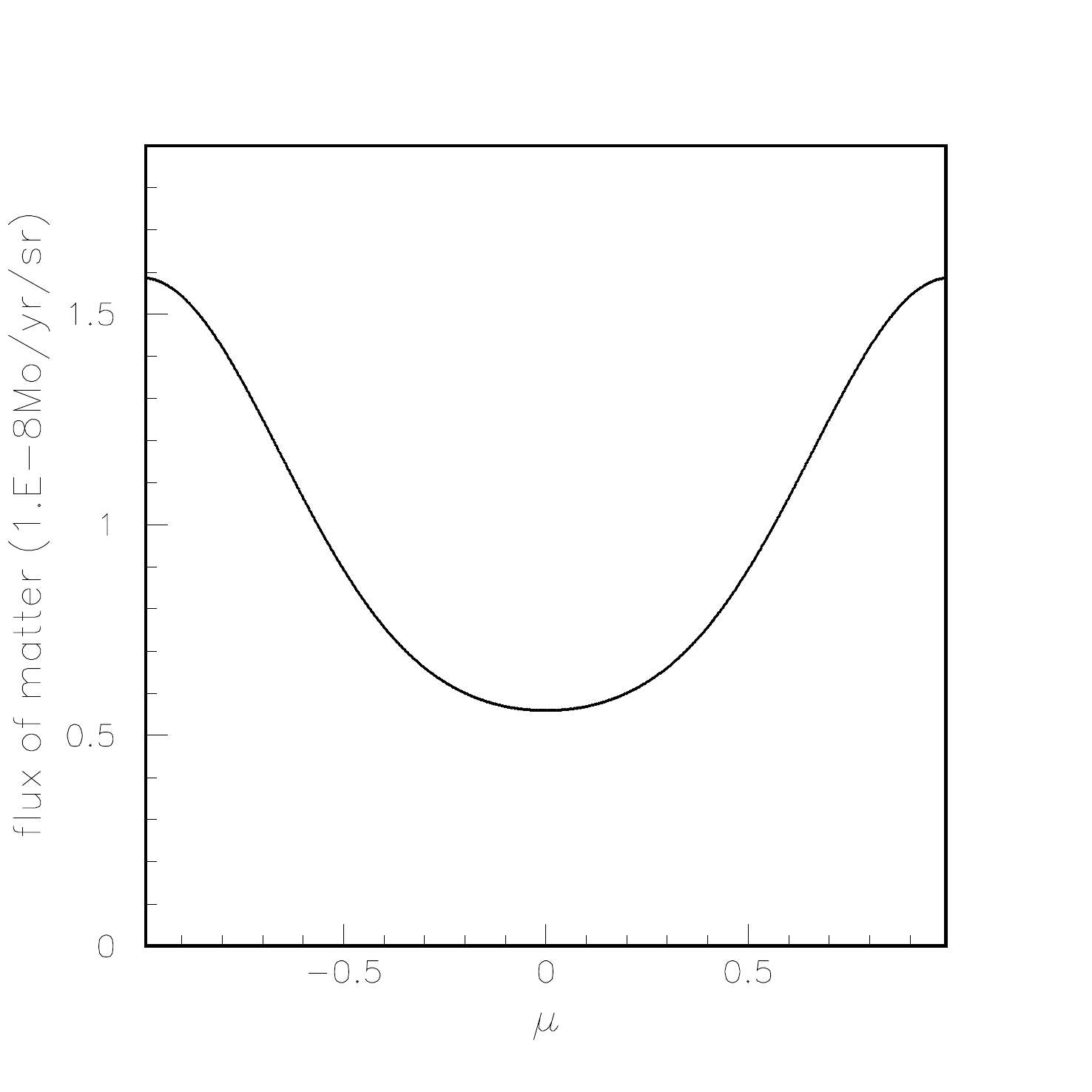,angle=0,width=8.0cm}
\caption{The flux of matter as a function of $\mu$.}
  \label{profmdot}
\end{figure}

\begin{figure}
\centering
\epsfig{figure=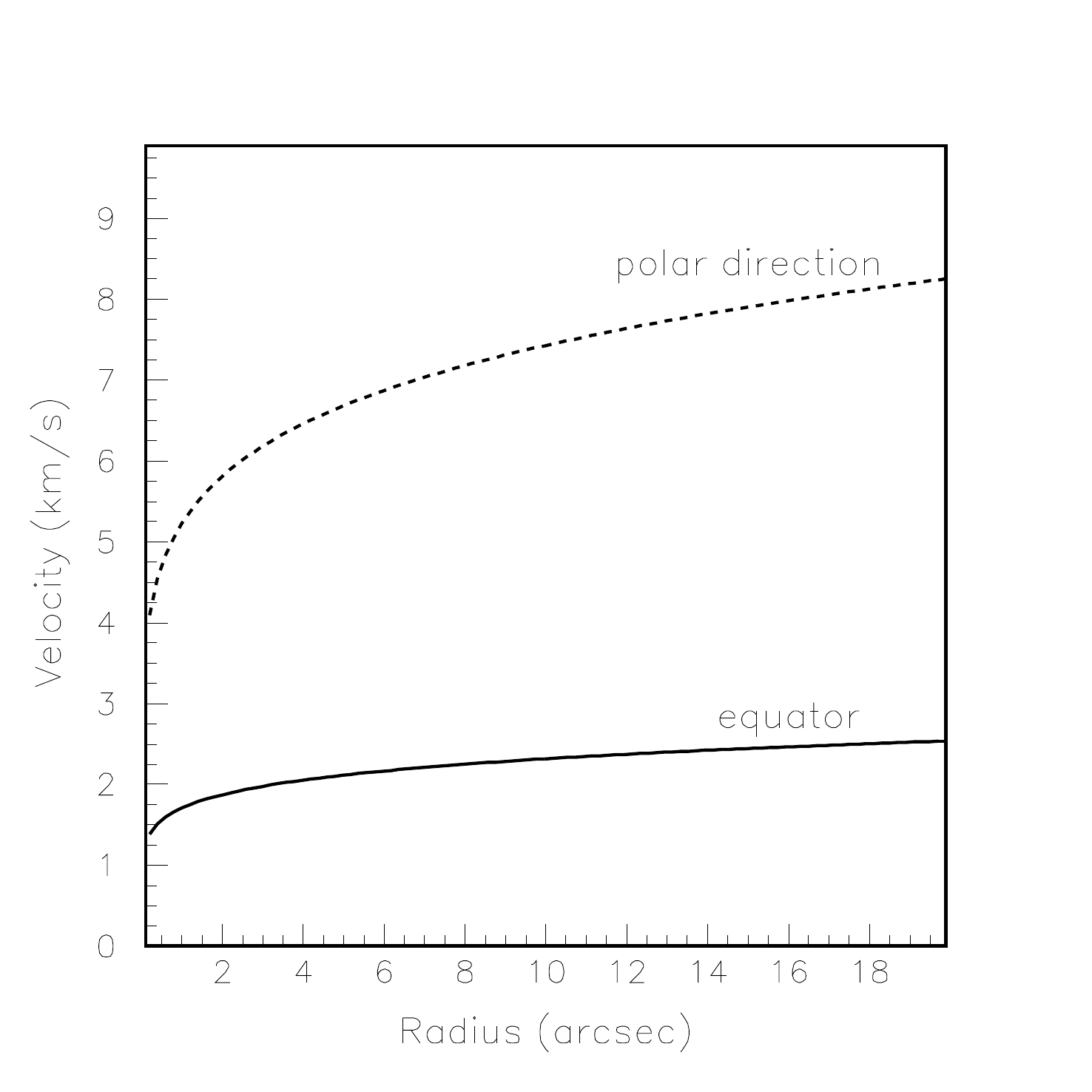,angle=0,width=8.0cm}
\caption{Velocity profiles in the equatorial plane ($\mu$ = 0) and along 
the polar axis ($\mu = \pm$ 1).}
  \label{velgradient}
\end{figure}

In our preferred model of RS\,Cnc, the density and the velocity are varying 
smoothly from the equatorial plane to the polar axis. 
The profiles of the density and the velocity are shown in Figs.~\ref{profd} 
and \ref{profv}, respectively. The latitude ($\theta$, $\mu$=sin$\theta$) 
dependence was obtained by a combination of exponential functions 
of $\mu$. In order to adjust 
the parameters of the model we used the {\mbox {\sc minuit}} 
package from the CERN program library (James \& Roos 1975), 
which minimizes the sum of the square of the deviations (i.e. modeled minus 
observed intensities). The minimization is obtained on the CO1-0 spectral map, 
which has the best quality, and the same parameters are applied for the 
\COtwo ~map. The flux of matter varies from 
0.53$\times10^{-8}$\,\Msold\,sr$^{-1}$ in the equatorial plane 
to 1.59$\times10^{-8}$ \Msold\,sr$^{-1}$ in the polar 
directions (Fig.\,\ref{profmdot}). The total mass loss rate is 
1.24$\times10^{-7}$ \Msold. The mass loss rate integrated within the two polar 
cones ($|\mu|>$0.5) is 0.83$\times10^{-7}$ \Msold. 
The exponent in the velocity profile varies from $\alpha$\,=\,0.13 in the 
equatorial plane, to $\alpha$\,=\,0.16 in the polar directions 
(Fig.\,\ref{velgradient}).

\begin{figure*}
\centering
\epsfig{figure=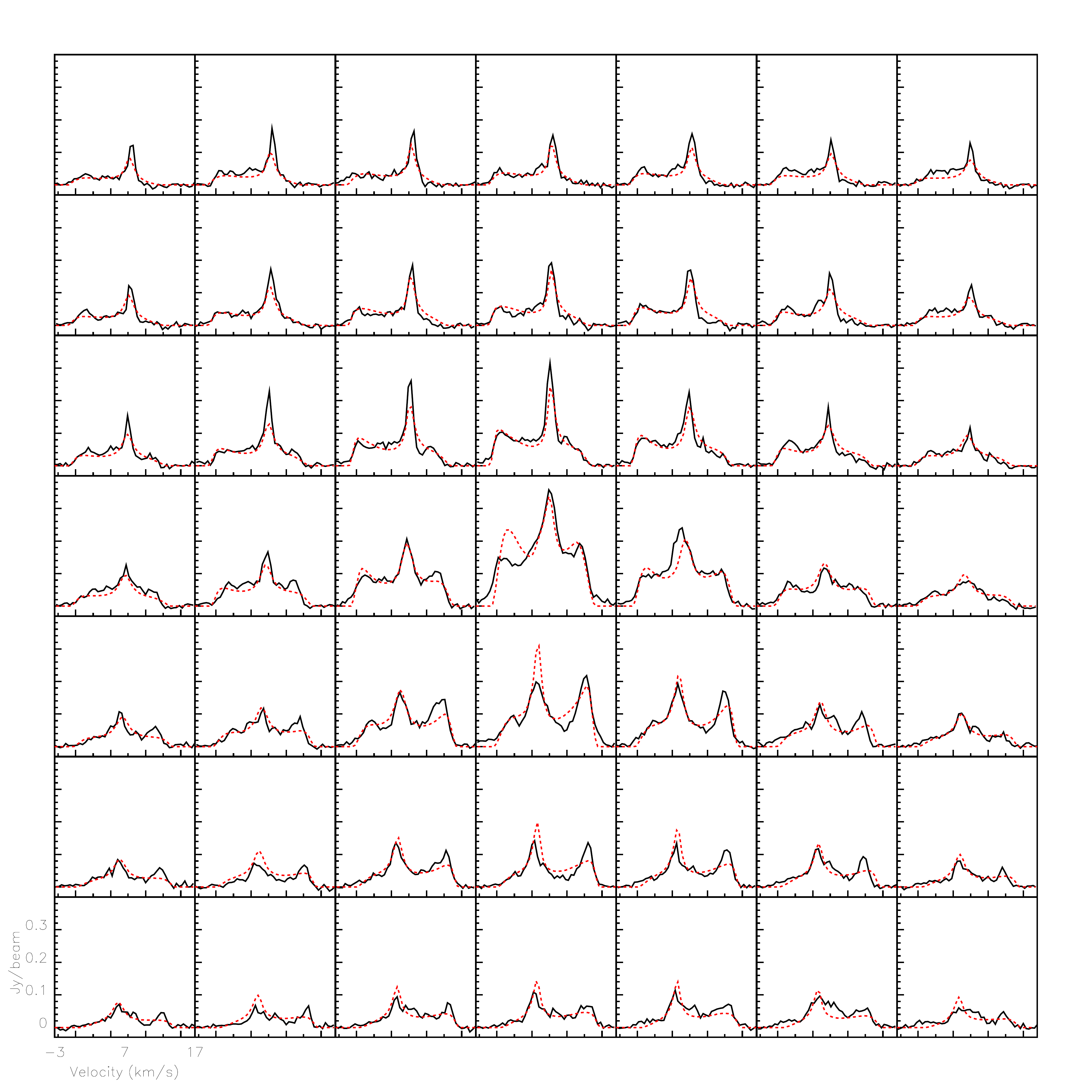,angle=0,width=9.4cm}
\caption{Central part of the CO1-0 spectral map of RS Cnc together with 
the fits (dotted lines, in red in the electronic version) obtained in 
Sect.~\ref{application}.}
  \label{FitdataCO10}
\end{figure*}

\begin{figure*}
\centering
\epsfig{figure=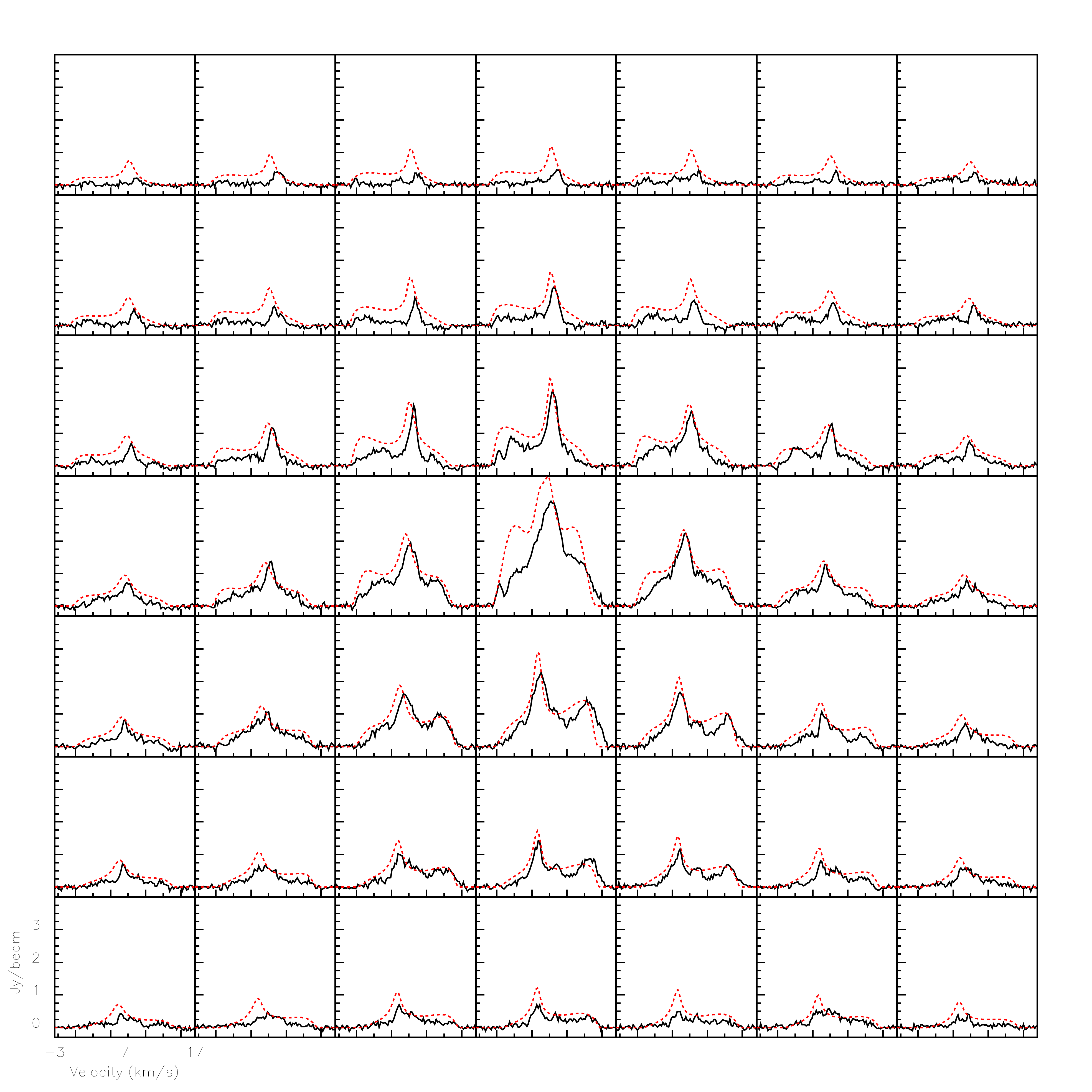,angle=0,width=9.4cm}
\caption{Same as in Fig.~\ref{FitdataCO10}, but for CO2-1. The original 
Libert et al. (2010) data have been resampled to 0.2 \kms.}
  \label{FitdataCO21}
\end{figure*}

In Figs.~\ref{FitdataCO10} and \ref{FitdataCO21}, we present a comparison of 
the spectra obtained in CO1-0 and CO2-1 together with the results of the 
model. We obtain a good compromise between the 2-1 and 1-0 data and the model, 
although with a slight excess of the model in 2-1 in particular in the central 
part of the map. The orientation of the source obtained with the minimization 
algorithm is defined by AI\,=\,52$^{\circ}$ (angle of inclination 
of the polar axis over the plane of the sky), and PA\,=\,10$^{\circ}$ 
(position angle of the projection of this axis over the plane 
of the sky). 

\section[]{VLA and JVLA Observations}
\label{HIdata}
\HI imaging observations of RS~Cnc obtained with the Very Large Array
(VLA) \footnote{The VLA of the National Radio Astronomy
  Observatory (NRAO) is operated by Associated
  Universities, Inc. under cooperative agreement with the National
  Science Foundation.} 
in its D configuration (0.035-1.03~km baselines) were previously presented
by Matthews \& Reid (2007; see also Libert et al. 2010). Those data were 
acquired using dual circular polarizations and a 0.77~MHz bandpass. 
On-line Hanning smoothing was applied in the VLA correlator, yielding 
a data set with 127 spectral channels and a channel spacing of 6.1~kHz
($\sim$1.29~\kms). Further details can be found in Matthews \& Reid.

For the present analysis, the VLA D configuration data were combined with new 
\HI 21-cm line observations of RS~Cnc obtained using the Jansky Very Large 
Array (JVLA; Perley et al. 2011) in its C configuration (0.035-3.4~km 
baselines). The motivation for the new observations was to obtain information 
on the structure and kinematics of the \HI emission on finer spatial
scales than afforded by the D configuration data alone, thereby enabling 
a more detailed comparison between the \HI and CO emission (see 
Sect.~\ref{COdata}).

The JVLA C configuration observations of RS~Cnc were obtained during 
observing sessions on 2012 March~2 and 2012 April~19. A total of 4.8~hours was
spent on-source. Observations of RS~Cnc were interspersed with
observations of the phase calibrator J0854+2006 approximately every 20 minutes.
3C286 (1331+305) was observed as a bandpass and absolute flux calibrator.

The JVLA WIDAR correlator was configured with 8 subbands across each of two 
independent basebands, both of which measured dual circular
polarizations. Only data from the first baseband pair (A0/C0) were
used for the present analysis. Each subband had a bandwidth of
0.25~MHz with 128 spectral channels, providing a channel spacing of 
1.95~kHz ($\sim$0.41~\kms). The 8 subbands were tuned to
contiguously cover a total bandwidth of 2~MHz.

The bulk of the JVLA data were taken with the central baseband frequency
slightly offset from the LSR velocity of the star. However, additional 
observations of the phase and bandpass calibrators were made with the 
frequency center shifted by $-1.5$~MHz and $+$1.5~MHz, respectively,  
to eliminate contamination from Galactic \HI emission in the band and 
thus permit a robust bandpass calibration and more accurate bootstrapping of
the flux density scale.

Data processing was performed using the Astronomical Image Processing
System (AIPS; Greisen 2003). Data were loaded into AIPS directly from archival
science data model (ASDM) format files using the {\sc BDFI}n program available 
in the Obit software package (Cotton 2008). This permitted the creation of
tables containing on-line flags and system power measurements.

After updating the antenna positions and flagging corrupted data, 
an initial calibration of the visibility data  was performed using the AIPS 
task {\sc TYAPL}, which makes use of the system power measurements to provide
optimized data weights (Perley 2010). Calibration of the bandpass and  
the frequency-independent portion of the complex gains was subsequently 
performed using standard techniques, taking into account the special
considerations for JVLA data detailed in Appendix~E of the 
AIPS Cookbook.\footnote{http://www.aips.nrao.edu/cook.html} The gain solutions 
for the fifth subband were interpolated from the adjacent subbands because 
of line contamination. Following these steps, time-dependent frequency
shifts were applied to the data to compensate for the Earth's motion.
\begin{figure*}
\centering
\epsfig{figure=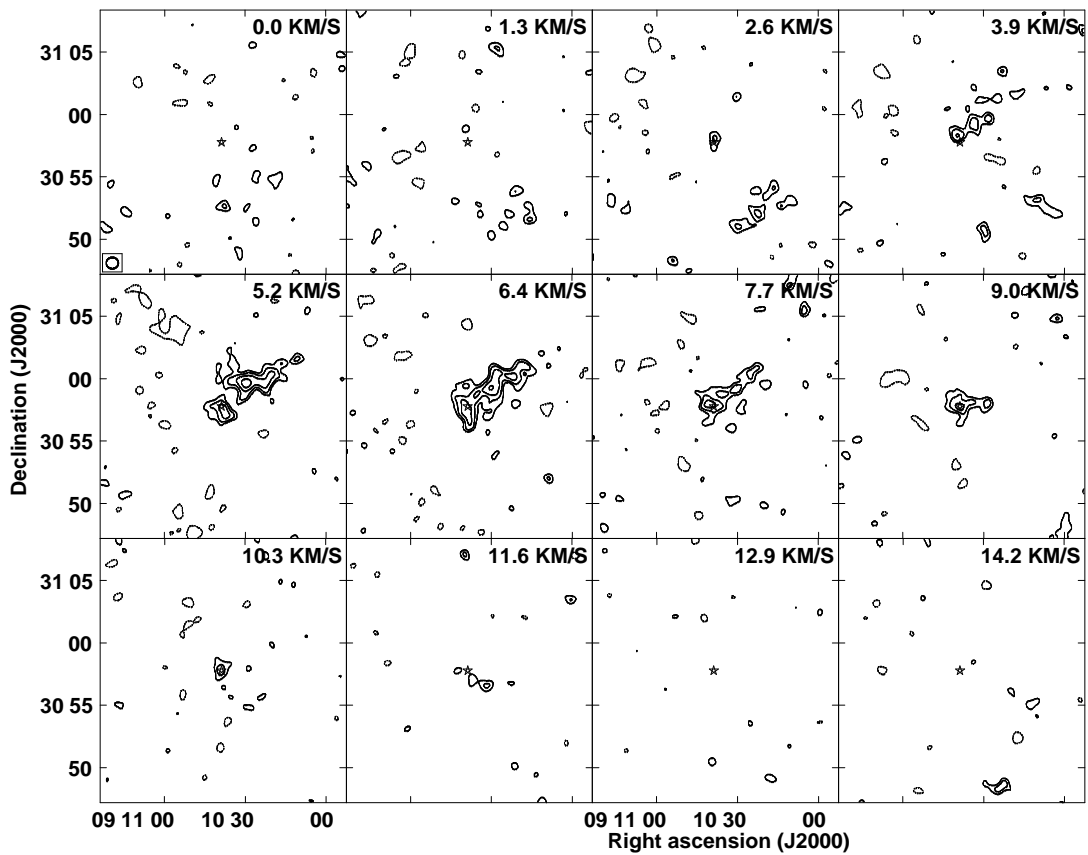,angle=0,width=17.0cm}
\caption{\HI channel images of RS~Cnc obtained from combined (J)VLA C and D
  configuration data. Contour levels are
  ($-$6[absent],$-$4.2,$-$3,3,4.2,6,8.5)$\times$1.4~mJy beam$^{-1}$. 
  A $u$-$v$ tapering of 4k$\lambda$ (see Sect.~\ref{HIimaging}) was
  used to produce these images.  The
  lowest contour is $\sim 3\sigma$. The synthesized beam size is
  \as{57}{9}$\times$\as{54}{2}. The star symbol marks the stellar 
position of RS\,Cnc from Hipparcos.\protect\label{fig:HIcmaps}}
\end{figure*}

\subsection{Imaging the Continuum}
An image of the 21-cm continuum emission within a $\sim50'$ region centered 
on the position of RS~Cnc was produced using only the C configuration data. 
After excluding the first and last two channels of each
subband and the portion of the band containing line emission, the
effective bandwidth was $\sim$1.7~MHz in two polarizations.  

Robust +1 weighting (as implemented in AIPS) was used to create the continuum 
image, producing a synthesized beam of \as{17}{4}$\times$\as{14}{3}. 
The RMS noise in the resulting image was $\sim$\,0.11\,mJy\,beam$^{-1}$. 

The brightest continuum source within the $\sim30'$ JVLA primary beam
was located $\sim$\am{12}{6} northwest of RS~Cnc, with a flux density of
110$\pm$1~mJy. No continuum emission was detected from RS~Cnc itself, 
and we place a 3$\sigma$ upper limit on the 21~cm continuum emission at 
the position of the star within a single synthesized beam to be $<$0.33~mJy. 
Using Gaussian fits, we compared the measured flux densities of the six 
brightest sources in the primary beam with those measured from NRAO VLA Sky 
Survey (Condon et al. 1998), and found the respective flux densities to be 
consistent to within formal measurement uncertainties.

\subsection{Imaging the \HI Line Emission\protect\label{HIimaging}}
Because the C and D configuration observations of RS~Cnc were obtained with 
different spectral resolutions, prior to combining them the C configuration 
data were boxcar smoothed and then resampled to match the channel spacing 
of VLA D configuration data. Subsequently, the C and D configuration 
data sets were combined using weights that reflected their respective 
gridding weight sums as reported by the
AIPS task {\sc IMAGR}. The combined data set contained 127 spectral
channels with a channel spacing of 6.1~kHz and spanned the LSR velocity
range from $+81.2$~\kms\ to $-$81.2~\kms. 
Before imaging the line emission, the continuum was subtracted from the
combined data set using a first order fit to the real and imaginary
parts of the visibilities in the line-free portions of the band (taken
to be spectral channels 10-50 and 77-188).

Three different \HI spectral line image cubes were produced for the
present analysis. The first used
natural weighting of the visibilities, resulting in a synthesized beam
size of \as{36}{2}$\times$\as{31}{6} at a position angle (PA) of $87^{\circ}$ 
and an RMS noise of $\sim$1.1~mJy beam$^{-1}$ per channel. For the
second image cube, the visibility data were tapered using 
a Gaussian function with a width at the 30\% level of 4~k$\lambda$ in
the $u$ and $v$ directions. This resulted in a synthesized beam of 
\as{57}{9}$\times$\as{54}{1} at PA=$-88^{\circ}$ and an RMS noise level of
$\sim$1.4~mJy beam$^{-1}$ per channel. The third data cube
used Gaussian tapering of 6~k$\lambda$, resulting in a synthesized
beam of \as{49}{5}$\times$\as{44}{8} at PA=89$^{\circ}$ and RMS noise
of $\sim$1.3~mJy beam$^{-1}$ per channel.

\section{\HI Imaging Results}
\subsection{The Morphology of the \HI Emission}
\begin{figure}
\centering
\epsfig{figure=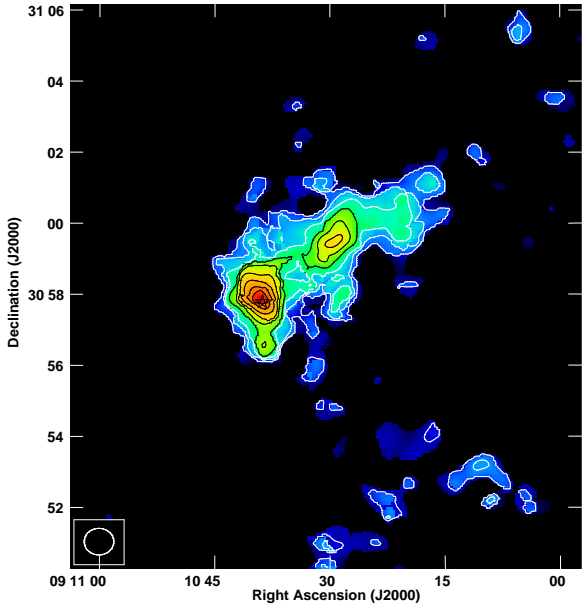,angle=0,width=9.0cm}
\caption{\HI total intensity map, derived from data with 6~k$\lambda$
  tapering (see Sect.~\ref{HIimaging}). Contour levels are
  (1,2,3...9)$\times$6.6~Jy beam$^{-1}$ m s$^{-1}$. 
This image was derived by summing the emission over the
velocity range from 2.6 to 11.6~\kms; to minimize the noise
contribution to the map, data that did not exceed a 2$\sigma$
threshold after smoothing the data spatially and spectrally by a
factor of 3 were blanked. The star symbol marks the stellar 
position of RS\,Cnc. \protect\label{fig:HImom0}}
\end{figure}

Figure~\ref{fig:HIcmaps} shows \HI channel images obtained from the
combined (J)VLA  C+D configuration data. Statistically significant
emission ($\ge 4\sigma$) is detected at or near the stellar position over 
the range of  LSR velocities from 2.6 to 11.6~\kms. The bulk of this emission 
appears to be associated with the circumstellar wake of RS~Cnc. 
This can be seen even more clearly in Fig.~\ref{fig:HImom0}, 
where we present an \HI total intensity map of
RS~Cnc, derived by summing the emission over the above velocity range.

As previously reported by Matthews \& Reid (2007), the \HI associated with 
RS~Cnc  comprises two components: a compact structure centered close to the
stellar position (the ``head''), of size 75$''\times$40$''$, 
and an extended wake of material that is 
known to trail directly opposite the direction of space motion of the
star (the ``tail''). The tail has a measured extent of $\sim 6'$
($\sim$0.25~pc) in the plane of the sky.

Based on Fig.~\ref{fig:HImom0} we find the peak \HI column density within 
the ``head'' of RS~Cnc, $\sim16''$ north of RS~Cnc (i.e., offset from the 
stellar position by roughly half a synthesized beam).

Our new \HI data clearly confirm the previous suggestions that the \HI 
emission surrounding the position of RS~Cnc is elongated and that the position 
angle of this elongation is consistent with the CO outflow described above 
(see Matthews \& Reid 2007; Libert et al. 2010).  This is evident in the 
\HI total intensity map (Fig.~\ref{fig:HImom0}) and in the channel image 
centered at $V_{\rm LSR}=6.4$~\kms\ (Fig.~\ref{fig:HIcmaps}). 
Fig.~\ref{fig:HImom0} also shows evidence for two lobes of emission, extending 
north and south respectively from the more compact ``head'' of the \HI emission
structure. Each of these lobes extends to $\sim100''$ from the stellar
position. The correspondence between the elongation of the \HI
emission in RS~Cnc's head and the position angle of the molecular
outflow traced in CO immediately suggest the possibility that the \HI 
is tracing an extension of the molecular outflow beyond the molecular
dissociation radius. An examination of the gas kinematics seems to
reaffirm this picture (see Sect.~\ref{HIkinematics}).

\begin{figure}
\centering
\epsfig{figure=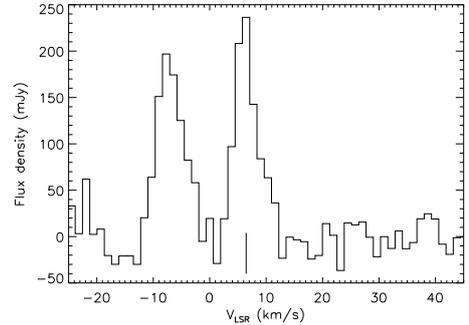,angle=0,width=8.0cm}
\caption{Spatially integrated 
\HI spectrum of RS~Cnc derived from naturally-weighted 
(J)VLA C+D configuration
data. The vertical bar indicates the stellar systemic velocity derived
from CO observations. The spectral feature centered near the stellar
systemic velocity is circumstellar in origin, while the second peak
near $V_{\rm LSR}=-7$~\kms\ results from interstellar contamination
within the measurement aperture. \protect\label{fig:HIspectrum}}
\end{figure}

\subsection{The Global \HI Spectrum and Total \HI Mass}
\label{globalHIspectrum}
An integrated \HI spectrum of RS~Cnc is shown in Fig.~\ref{fig:HIspectrum}. 
The narrow, roughly Gaussian shape of the line profile is typical of many 
of the other AGB stars detected in \HI (e.g., G\'erard \& Le Bertre 2006; 
Matthews et al. 2013), but it contrasts with the two-component CO line 
profile of RS~Cnc (Fig.~\ref{spectralmapCO10}). The \HI line centroid is 
also slightly offset from the stellar systemic velocity 
derived from the CO spectra (see Table~\ref{basicdata}),
and as already seen from Fig.~\ref{fig:HIcmaps}, the velocity range 
of the detected \HI emission is significantly smaller than that of the CO
emission. This suggests either that the atomic hydrogen has been slowed
down by its interaction with the surrounding medium and/or that the material 
we detect in \HI emission originated during an earlier epoch of mass loss 
during which the maximum outflow speeds were lower.

We have derived the velocity-integrated \HI flux density for RS~Cnc 
by integrating the emission in each spectral channel between 2.6-11.6~\kms\ 
within a \am{6}{8}$\times$\am{6}{4} rectangular aperture
centered at the middle of the tail  
($\alpha_{\rm J2000}=9^{\rm h}$ 10$^{\rm m}$ 29.8$^{\rm s}$, 
$\delta_{\rm J2000}=30^{\circ} 58'$ \as{47}{3}). 
We use the naturally weighted data cube for this
measurement, after correcting for the attenuation of the primary
beam. Using this approach, we measure an integrated \HI flux density
$\int S_{\rm HI} dV= 1.14\pm$0.03~Jy~\kms. At our adopted distance to RS~Cnc, 
this translates to an \HI mass of 5.5$\times10^{-3}~M_{\odot}$,  
where we have used the standard relation $M_{\rm HI} = 
2.36\times10^{-7}d^{2} \int S_{\rm HI} dV$.
Here $d$ is the distance in parsecs, $V$ is the velocity in \kms, 
and the units of $M_{\rm HI}$ are solar masses. 

The integrated \HI flux density that we derive for RS~Cnc is a factor
of $\sim$2.5 times higher than reported previously by Matthews \& Reid
(2007). We attribute this difference to a combination of three factors. First,
the improved sensitivity and short spacing $u$-$v$ coverage of our combined 
C+D configuration data improves our ability to recover weak, extended
emission. Secondly, Matthews \& Reid measured the integrated emission
within a series of irregularly shaped ``blotches'' in each spectral channel, 
defined by their outer 2$\sigma$ contours. While this approach minimizes 
the noise contribution to each measurement, it can also exclude weak, 
extended emission or noncontiguous emission features from the sum. Lastly, 
Matthews \& Reid measured their integrated flux density from a tapered
image cube. Based on the original VLA data alone, 
we found that for our presently adopted measurement aperture, this 
results in an integrated flux density that is $\sim$ 40\% lower
compared with a measurement from a naturally
weighted data cube. Although our new \HI mass estimate is
higher than the previous value reported from VLA measurements, it is
significantly less than reported by Libert et al. (2010) from NRT
measurements ($M_{\rm HI}\approx0.04~M_{\odot}$). We now suspect that
the NRT measurement suffered from local confusion around 7 \kms. 

\begin{figure}
\centering
\epsfig{figure=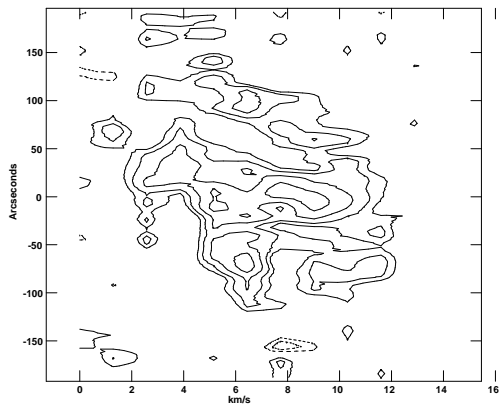,angle=0,width=8.0cm}
\caption{\HI position-velocity plot derived from naturally weighted
  data along PA=10$^{\circ}$ and centered
  on the position of RS~Cnc. Contour levels are
  ($-5.6$[absent],$-$4,$-$2.8,$-$2,2...5.6)$\times$1.1 mJy
  beam$^{-1}$. The lowest contour level is
  $\sim2\sigma$. The origin of the $y$-axis is the stellar 
position.\protect\label{fig:HIPV}}
\end{figure}

\subsection{Kinematics of the \HI Emission}
\label{HIkinematics}
One of the key advantages of \HI 21-cm line observations for the
study of circumstellar ejecta is that they provide valuable kinematic
information on material at large distances from the star. For example,
for the AGB stars Mira and X~Her, both of which have trailing \HI wakes 
analogous to RS~Cnc, Matthews et al. (2008) and Matthews et al. (2011),
respectively, measured systematic velocity gradients along the length 
of the circumstellar wakes and used this information to constrain the 
timescale of the stellar mass-loss history (see also Raga \& Cant\'o 2008).
Interestingly, RS~Cnc shows no clear evidence for a velocity gradient along
the length of its wake. This can be seen simply from inspection of the
channel maps in Fig.~~\ref{fig:HIcmaps}. RS~Cnc also contrasts with 
the other stars known to have trailing \HI wakes in that its space velocity 
is much lower, $\sim$15~\kms\ compared with $V_{\rm space}\gsim$57~\kms\ 
(see Matthews et al. 2013). 

While no velocity gradient is seen along the tail of RS~Cnc, the morphology 
of the tail gas reveals evidence for the importance of hydrodynamical effects. 
For example, a ``wavy'' structure is evident along the length of the tail in 
the channel image centered at $V_{\rm LSR}=6.4$~\kms\ 
(Fig.~\ref{fig:HIcmaps}). Additionally, the \HI total intensity image in
Fig.~\ref{fig:HImom0} shows a region of enhanced column density
approximately half way long the length of this tail. A narrow stream
of gas appears to connect this ``presque isle'' to the head of RS~Cnc.

To facilitate comparison between the kinematics of the \HI and CO emission 
in the CSE of RS~Cnc, we present in Fig.~\ref{fig:HIPV} an \HI 
position-velocity plot extracted along PA=10$^{\circ}$---i.e., the position 
angle of the CO outflow described above.  This plot shows that the kinematics 
of the atomic gas in the ``head'' of RS~Cnc are rather complex.

We see that the central region of the head ($r\lsim\pm50''$) contains
emission spanning from $\sim$2 to 12~\kms---i.e., the full velocity
range over which \HI has been detected. However, all of the gas
blueward of $\sim$4~\kms\ lies north of the stellar position. This is
suggestive of a relation to the high-velocity CO outflow, in which
blueshifted emission is seen north of the star. Further, the velocity
spread of the emission is smaller than seen in the molecular gas.

At distances of $>\pm 50''$ from RS~Cnc, the gas kinematics appear different 
on the northern and southern sides of the head. In the south, we see evidence 
of \HI emission near $V_{\rm LSR}\approx6$~\kms\ and 
$V_{\rm LSR}\approx10$~\kms\ extending to $\sim100''$ from the stellar 
position. These features may represent the atomic
counterparts to the south polar and equatorial molecular outflows, 
respectively. In the north, the high-latitude emission is dominated by 
a plume of gas whose velocity decreases from $\sim$9~\kms\ at
$\sim50''$ north to $\sim$6~\kms\ (close to the stellar systemic
velocity) at $\sim125''$ north. In Fig.~\ref{fig:HImom0} we see that the
morphology and position angle of the emission at this location are suggestive 
of this being the extended northern counterpart to high-speed polar CO 
outflow. Alternatively, it could be material stripped from the equatorial 
regions.

\section[]{Discussion}
\label{discussion}

\subsection{CO model}
\label{model}

In Sect.~\ref{application}, we have presented our preferred CO model, with 
a continuous density distribution from the equatorial plane to the poles. 
However, we tried several other configurations before selecting this model. 
Our first trial consisted in using directly the model proposed 
by Libert et al. (2010) of a bipolar flow and of an equatorial disk, 
separated by a gap, and with no velocity gradient in the outflows. 
The calculated spectra showed spikes at \Vstar\,$\pm$\,2\,\kms, and 
\Vstar\,$\pm$\,6\,\kms, corresponding to the velocities selected in the model
(projected to the line of sight) whereas the observations reveal much
smoother spectra. Also, 
the drifts in the velocities seen in the spectral maps were not reproduced. 

In a second series, we introduced a gradient in the velocities in order 
to reproduce these drifts, all other parameters 
being kept identical. The agreement was generally improved, except for the 
central spectra. We have thus been led to subdivide the equatorial disk, 
introducing an inner part ($\pm$\,15$^{\circ}$ of the equatorial plane) 
where the velocity is kept slow, and an outer part 
(from 30 to 45$^{\circ}$, and from --30 to --45$^{\circ}$) where the 
velocity is increased. Finally, as the gap between the equatorial disk and
the polar outflows does not seem physically justified, we introduced 
smooth functions of the latitude for describing the velocity and the density, 
and obtained the results presented in Sect.~\ref{application}. 
With this final step, the improvement on the sum of the square of the 
residuals was at least of a factor 2. 

\begin{figure*}
\centering
\epsfig{figure=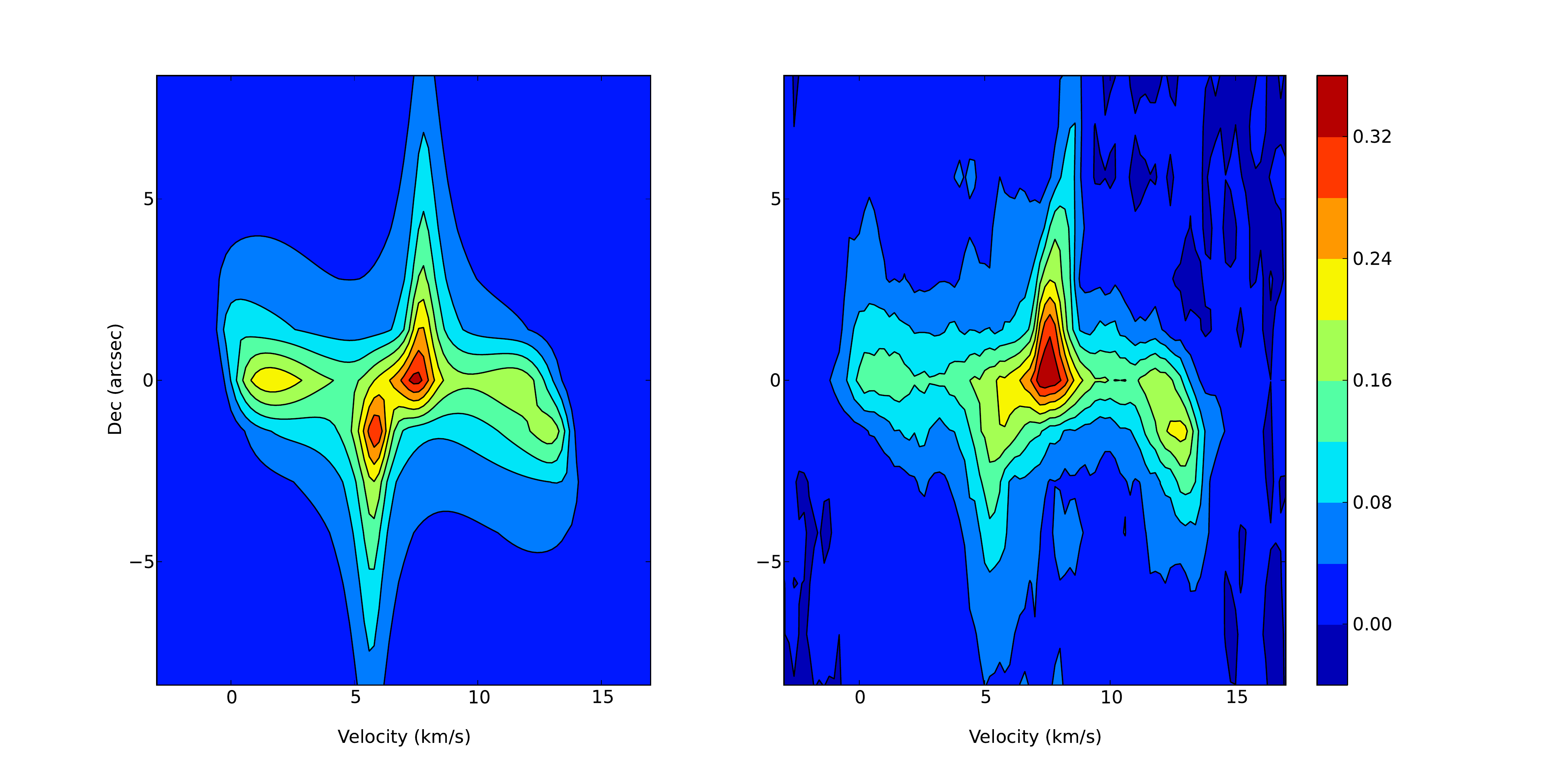,angle=0,width=14.0cm}
\caption{Left panel: synthetic position-velocity diagramme 
(offset in declination, 
from the 2000.0 Hipparcos position, against velocity) in CO1-0, 
obtained for our preferred model. The scale is in Jy/beam, and the channel 
width is 0.2 \kms. Right panel: 
observed position-velocity diagramme with same conditions.}
\label{synthposvel}
\end{figure*}

One of the conclusions of these exercises is that the notion of a disk 
as a separate entity is perhaps misleading. The data are consistent with 
a low velocity outflow that extends in latitude far from the equatorial plane. 
On the other hand, 
a high velocity outflow along the polar direction is clearly needed. 
Although we cannot assert that our model is a unique representation of the 
close environment of RS\,Cnc, it is the simplest that we could find and that 
gives an approximate reproduction of the spectral maps available in CO1-0 and 
2-1. The position-velocity diagrammes with two S-shaped features 
in opposition are often taken as evidence of disk/outflow structures 
(Nakashima 2005, Libert el al. 2010). In Fig.~\ref{synthposvel}, we present 
the position (Dec)-velocity diagramme in CO1-0 obtained with our preferred 
model. The two S-shaped features are reproduced with the correct sizes and 
velocity amplitudes. It  
illustrates that our model can account as well for such a kind of diagramme. 

The estimate of the mass loss rate that we obtain through our modeling 
agrees with that of Knapp et al. (1998, cf. Table\,\ref{basicdata}), 
but not with that of Libert et al. (2010, 
7.3\,10$^{-7}$\,\Msold ~at 143\,pc). The latter assumed an optically 
thick wind in CO1-0, which, from the present work, appears unlikely. 
As discussed in Winters et al. (2003), this hypothesis may lead to an 
overestimate of the mass loss rate by a factor $\sim$ 3.5. We investigated 
the effect of the optical depth in our model, and found that self-absorption 
becomes effective only in the central part of the circumstellar shell 
in CO2-1. Furthermore, the CO2-1 and 3-2 spatially integrated spectra 
obtained by Knapp et al. are fitted satisfactorily by our model. 

The simultaneous fit to the CO1-0, 2-1 and 3-2 data supports 
the temperature profiles adopted from Sch\"oier \& Olofsson (2010). However, 
we note that a steeper temperature profile would improve the 
quality of line-profile fits in the 2-1 spectral map, with almost no effect 
in 1-0. This is probably related to the fact that the 2-1 line starts to be 
optically thick for the lines of sight close to the central star. Clumpiness 
in the outflow could as well affect more the 2-1 data than the 1-0 ones. 

An intriguing feature of the modeling is the need of a velocity gradient 
inside the CO shell. It is needed in order to reproduce the shifts in 
velocity of the blue and red peaks with distance to the central star. 
In a stationary case, it means that the outflow is still accelerated 
at distances larger than a few hundred AU (few 10$^{15}$\,cm). 
This is surprising because in models of dust-driven winds a terminal velocity 
is reached at a distance of $\sim$ 20 stellar radii (few 10$^{14}$ cm, 
e.g. Winters et al. 2000). 
Other acceleration processes should probably be considered, for instance 
the radiation force on molecules (J{\o}rgensen \& Johnson 1992), 
which may play a role in low mass loss rate winds. 

On the other hand, evidence of accelerating outflows has been obtained 
in some bipolar Pre-Planetary Nebulae. Recently Sahai et al. (2013) find 
a velocity gradient of 4 \kms ~from $\sim$\,4$\times$10$^{16}$ to 10$^{17}$ cm 
in the 'waist' of the low-luminosity (300\,\Lsol) central star of 
the Boomerang Nebula (IRAS 12419--5414). 
Finally, although ad hoc, we cannot exclude that the flow is not stationary, 
and that we are witnessing a decrease with time of the expansion velocity.

In other CO-models, such as that of 
Sch{\"o}ier \& Olofsson (2001), the expansion velocity is assumed to be 
constant throughout the CO shell. It is noteworthy that in these models 
the goal is to reproduce the spatially integrated spectra, and that turbulence 
is invoked with a typical value of $\sim$ 0.5 \kms ~throughout the entire 
flow. In our modeling, we do not need to invoke turbulence.

\subsection{binarity}
\label{binarity}

The presence of Tc lines in its optical spectrum shows that RS\,Cnc is 
evolving on the TP-AGB, and that it has already undergone several thermal 
pulses and dredge-up events. Busso \& Trippella (2013, personal 
communication), using recent prescriptions for mass loss (Cristallo et al. 
2011) and the revision of the s-process element production by Maiorca et al. 
(2012), fitted the abundances and $^{12}$C/$^{13}$C ratio determined by Smith 
\& Lambert (1986) with a stellar  evolution model of a 1.6 \Msol ~star 
in its 4th dredge-up episode. \rscnc ~is clearly an intrinsic S-type star 
that does not owe its peculiar abundances to a mass transfer from a more 
evolved companion (Van Eck \& Jorissen 1999). 

\begin{figure}
\centering
\epsfig{figure=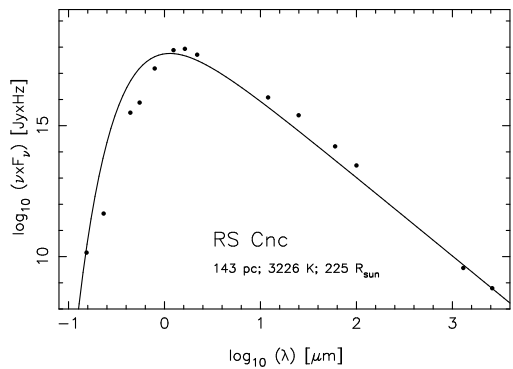,angle=0,width=8.0cm}
\caption{Energy distribution of \rscnc. The continuous line corresponds to 
a blackbody with size and temperature determined by Dumm \& Schild (1998, 
see also Table~\ref{basicdata}).}
  \label{specRSCnc}
\end{figure}

The energy distribution of \rscnc ~is presented in Fig.~\ref{specRSCnc}. 
It combines UV data from GALEX (at $\sim$ 154 and 232 nm), optical data from 
Mermilliod (1986), near-infrared data from 2MASS, far-infrared data from IRAS, 
and mm data from our work. There is no clear evidence of an UV excess that 
might reveal the presence of a warm companion. Also, we have found no evidence 
of a companion in the continuum map (Sect.~\ref{continuum}).  
The CO structure at 6.6 \kms ~(see Fig.~\ref{6p6kmpersec-channelmap10}) may 
hint to the presence of a companion that might be surrounded by a disk 
accreting material from the wind that would be seen only in a narrow range 
of velocities. In that case we would have a less evolved star, probably still 
on the main sequence, for instance a dwarf of less than 1.6\,\Msol. 
However, presently, we cannot discriminate between a clump in the outflow and 
a cloud around a companion. 

In the course of the modeling process described in Sect.~\ref{model}, 
we also introduced a disk in Keplerian rotation around the
central star. There was no improvement of the quality of the fits and we
did not include the rotating disk in our preferred model. 
It means that, presently, there is no evidence {\underbar {in the data}} 
of such a structure. However, we cannot exclude that new data with a better 
spatial resolution ($\leq$~1$''$) would bring it into evidence. 
Such a structure has been invoked for X Her, another semi-regular AGB star 
with composite CO line-profiles, by Nakashima (2005), although it could 
not be confirmed by Castro-Carrizo et al. (2010). 

The absence of evidence for a companion or a disk is puzzling. 
Another possibility for explaining the axi-symmetry in the outflow might be 
the rotation of the central star. Evidence for rotating cores in giant 
stars has recently been obtained by the satellites CoRoT and Kepler (e.g. 
Beck et al. 2012). These rotating cores might have an influence on the late 
mass loss of evolved stars when most of the stellar envelope has been removed. 
However, hydrodynamical models of stellar winds from rotating AGB stars 
predict a higher mass loss rate in the equatorial 
plane than in the polar directions (Dorfi \& H{\"o}fner 1996, 
Reimers et al. 2000), which does not agree 
with our finding for RS Cnc (see Fig.~\ref{profmdot}). Finally, 
magnetic fields have been invoked for producing equatorial disks 
in AGB stars (e.g. Matt et al. 2010). These models result also in higher
densities in the equatorial plane than in the polar directions,
something that we do not observe. 

\subsection{'head' and 'tail' in \HI}
\label{headtail}

The 'head-tail' structure revealed by the first observations of Matthews 
\& Reid (2007) is reaffirmed. The global shape agrees with the 
triangular-shaped image obtained by Spitzer at 70\,$\mu$m (Geise 2011), 
although the length of the IR tail ($\sim 4'$) 
appears smaller than that in H\,{\sc {i}}. This might be an effect of the 
temperature of the dust which is decreasing with distance to the central star, 
whereas the \HI emission should not depend on temperature, and is thus 
a better tracer of morphology. Matthews et al. (2013) have classified 
the structures observed in \HI around evolved stars in three categories. 
RS Cnc clearly belongs to Category 1 of {\it ``extended wakes, trailing the
motions of the stars through space''}, and resulting from the dynamical 
interaction between stellar winds and their local ISM. 

Our new JVLA data show additional detail that makes RS\,Cnc the best observed 
case. The 'head' is elongated in a direction consistent with the polar 
direction determined by the CO modeling. It corresponds to the direction 
in which the mass loss rate is maximum. However the velocity in 
\HI is clearly smaller than in CO, which means that the outflow has been 
slowed down. Nevertheless, there is presently no observational evidence 
for a termination shock that would mark the external limit of the freely 
expanding wind (Libert et al. 2007). If it really exists it must be located 
between 20$''$ (or 4\,10$^{16}$\,cm, the external limit of the CO wind), 
and 30$''$ (our spatial resolution at 21 cm). 

The new data show also sub-structures in the 'tail' that seem to be 
distributed in two groups, at $\sim 2.5'$ and 5$'$ from the star, 
that share the same elongated global shape as 
that of the 'head'. They could trace past episodes of enhanced mass loss. 
However the hydrodynamical models of the interaction between  stellar 
outflows from evolved stars and the ISM tend to show that the wind variations 
do not remain recorded in the density or velocity structure of the gas 
(e.g. Villaver et al. 2002, 2012). Also the sub-structures observed in the 
'tail' are evocative of the vortices suggested by numerical simulations 
(Wareing et al. 2007). The presence of a latitude dependence of the stellar 
outflow may help to develop such sub-structures (Raga et al. 2008).

A particularity of RS Cnc 'tail', as compared to the other objects in Category 
1, is that there is no observed velocity gradient. It means that the tail of 
RS Cnc lies in the plane of the sky. This could be an effect of the small 
value of the stellar radial velocity. It also suggests that the local ISM 
shares the same radial velocity as RS Cnc. This hypothesis is supported by 
the ISM confusion which peaks at $\sim$ 7\,\kms ~(cf. figures 11 to 13 in 
Libert et al. 2010, and Sect.~\ref{globalHIspectrum}). If this is correct 
it would mean that the velocity of RS Cnc relative to its surrounding 
medium is still smaller than its space velocity (15\,\kms). 

The mass in atomic hydrogen is now estimated at 0.0055 \Msol. 
Assuming 10\% of the matter in He, it translates to 0.008 \Msol. 
This does not account for a possible component of the 'head-tail' in molecular 
hydrogen and/or ionized hydrogen. Given the effective temperature of RS Cnc 
(\Teff = 3226\,K, Table~\ref{basicdata}), the presence of the former 
is unlikely (Glassgold \& Huggins 1983). 

Adopting the present mass loss rate estimated from the CO modeling, 
1.24$\times$10$^{-7}$ \Msold, we obtain a timescale of 64$\times$10$^3$\,years 
for the formation of the 'head-tail' structure. This estimate should be used 
with caution, as we observe structures in the tail that might be due to 
episodes of enhanced mass loss. Also, models of stellar evolution generally 
predict a mass loss rate increasing with time. 

The proper motions, corrected for solar motion (8 mas/yr in RA and 
17 mas/yr in Dec), would translate to a crossing time of 
$\sim 2\times10^4$ years for a 6$'$ structure, a factor 3 less than 
the previous estimate. 
Therefore, the matter in the tail appears to follow the star in its 
motion through space, an effect that has already been observed in the other 
stars of Category 1 (e.g. Mira, Matthews et al. 2008).

Finally, the narrow linewidth ($\sim$\,4\,\kms, Fig.~\ref{fig:HIspectrum}) 
that is observed in \HI can be used to constrain the temperature of the gas 
in the tail of RS Cnc. Assuming a Maxwellian distribution of the velocities 
and an optically thin emission, an \HI line shows a Gaussian profile with a 
FWHM\,=\,0.214$\times$T$^{1/2}$. We can thus estimate the average temperature 
in the tail at $\lsim$~350\,K. As discussed above there is no global 
kinematics broadening of the \HI line, because the tail lies in the plane 
of the sky, but there might be a contribution from vortices. 

\subsection{the multi-scale environment of RS\,Cnc}
\label{multiscale}

It is now possible to describe the RS Cnc outflow from $\sim$ 100 AU to 
its interface with the ISM at distances of tens of thousands of AU from the 
star. In the central part, the outflow exhibits an axi-symmetric morphology 
centered on the red giant. This structure develops on a scale $\geq$ 2000 AU, 
as probed by the CO lines. Our modeling shows that the flow is faster and
more massive along the polar directions than in the equatorial plane. 
However, our present spatial resolution does not allow 
us to uncover the physical mechanism that shape the outflow close to the mass 
losing star, nor to evaluate the nature and the role of the sub-structure 
observed at 1$''$ northwest around 6.6 \kms. 

The \HI line at 21\,cm enables us to trace the outflow beyond the molecular 
photodissociation radius. The flow is observed to be slowed down, and  
to keep the same preferential orientation, as observed in CO, 
over a distance of $\sim$ 8000 AU ($\sim$\,0.04\,pc). 
Further away (from 0.04 pc to  0.25 pc) the flow is distorted 
by the motion relative to the local ISM. It takes the shape of a tail which 
is narrowing down stream, and shows sub-structures ($\sim$\,0.01\,pc) that 
can be interpreted as vortices resulting from instabilities initiated 
at the level of a bow shock (Wareing et al. 2007). However the present 
spatial resolution is insufficient to observe the transition between the 
free-flowing stellar wind and the slowed-down flow in interaction with 
the ISM. Also, we do not observe directly the bow shock, perhaps because 
hydrogen from the ISM is ionized when crossing this interface 
(Libert et al. 2008). 

Recent hydrodynamic models predict that tails similar to that observed around 
RS Cnc may develop in the early phase of mass loss from AGB stars (Villaver et 
al. 2003, 2012). However, they seem to be found preferentially in stars moving 
rapidly through the ISM ($\geq$ 50 \kms). The axi-symmetry of the central flow 
might also play a role in the development of red giant tails, as other stars 
which have a composite line-profile in CO show also an H\,{\sc {i}}-tail 
(Matthews et al. 2008, 2011). The proximity of RS Cnc, the favorable observing 
conditions, and the well documented parameters of the system should allow 
us to perform a detailed comparison between observations and models.

\section[]{Conclusions}
\label{conclusion}

New PdBI CO1-0 and JVLA \HI interferometric data with higher spatial 
resolution have been obtained on the mass-losing semi-regular variable 
RS\,Cnc. These data combined with previous ones allow us 
to study with unprecedented detail the flow of gas from the central star 
to a distance of $\sim 10^{18}$\,cm. 

The previous CO observations of RS\,Cnc have been interpreted with a model 
consisting of an equatorial disk and a bipolar outflow. However, 
we find now that a better fit to the data can be obtained by invoking 
continuous axi-symmetric distributions of the density and the velocity, 
in which matter is flowing faster along the polar axis than in the 
equatorial plane. The polar axis is oriented at a position angle (PA) of 
10$^{\circ}$, and an inclination angle (from the plane of the sky) of 
52$^{\circ}$. It is aligned on the central AGB star. 
The mass loss rate is $\sim 1.24\times10^{-7}$ \Msold, 
with a flux of matter larger in the polar directions than in the equatorial 
plane. The axi-symmetry appears mainly in the velocity field and not in the 
density distribution. This probably implies that both stellar rotation 
and magnetic field are not the cause of the axi-symmetry. 
We also find that an acceleration of the outflow is possibly still 
at play at distances as large as 2$\times 10^{16}$\,cm from the central star.

The \HI data obtained at 21 cm with the (J)VLA show a 'head-tail' morphology. 
The 'head' is elongated in a direction consistent with the polar axis observed 
in the CO lines. 
The emission peaks close to the star with, at the present stage, 
no direct evidence of a termination shock. The 6$'$-long tail is oriented 
at a PA of 305$^{\circ}$, consistent with the proper motion of the star. 
It is resolved in several clumps that might develop from hydrodynamic 
effects linked to the interaction with the local interstellar medium. 
We derive a mass of atomic hydrogen of $\sim$ 0.0055 \Msol, and 
the timescale for the formation of the tail $\sim$ 64$\times$ 10$^3$ years, 
or more.

\begin{acknowledgements}
We acknowledge fruitful discussions with Pierre Darriulat, and are grateful 
for his continuous support. We are grateful 
to Maurizio Busso and to Oscar Trippella for their new determination of the 
evolutionary status of RS Cnc. We are grateful also to Hans Olofsson and 
to Fredrik Sch\"oier for providing their temperature structures of AGB 
circumstellar shells. We thank Eric Greisen for updates to AIPS tasks that 
were used for this work. We thank the LIA FVPPL, the PCMI, and the ASA 
(CNRS) for financial support. LDM gratefully acknowledges financial support 
from the National Science Foundation through award AST-1310930. 
Financial and/or material support from the Institute for Nuclear Science and 
Technology, Vietnam National Foundation for Science and Technology Development 
(NAFOSTED) under grant number 103.08-2012.34 and World Laboratory 
is gratefully acknowledged. The (J)VLA data used for this project 
were obtained as part of programs AM798 and AM1126. 
This research has made use of the SIMBAD and ADS databases.
\end{acknowledgements}

\end{document}